\documentclass{article}

\usepackage{arxiv}
\usepackage{float}
\usepackage{multicol}
\usepackage{multirow}
\usepackage{esint}

\usepackage{amsmath} 
\usepackage{amssymb}
\usepackage{calc}

\usepackage[utf8]{inputenc} 
\usepackage[T1]{fontenc}    
\usepackage{hyperref}       
\usepackage{url}            
\usepackage{booktabs}       
\usepackage{amsfonts}       
\usepackage{nicefrac}       
\usepackage{microtype}      
\usepackage[labelfont = bf]{caption}
\usepackage[labelfont = bf]{subcaption}

\newcommand{\volume}{%
  \makebox[0pt][c]{%
    \ooalign{$V$\cr\raisebox{0.15em}{\kern-0.1ex---}\cr}%
  }%
}

\usepackage{graphicx}
\usepackage[
backend=biber,
style=nature,
]{biblatex}
\usepackage{doi}
\title{C-ShipGen: A Conditional Guided Diffusion Model for Parametric Ship Hull Design}

\author{ \href{https://orcid.org/0000-0001-9893-8619}{\includegraphics[scale=0.06]{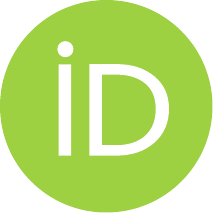}\hspace{1mm}Noah J.~Bagazinski}\thanks{Corresponding Author.} \\
	Department of Mechanical Engineering\\
	Massachusetts Institute of Technology\\
	Cambridge, MA 02139 \\
	\texttt{noahbagz@mit.edu} \\
	\And
	\href{https://orcid.org/0000-0002-5227-2628}{\includegraphics[scale=0.06]{orcid.pdf}\hspace{1mm}Faez ~Ahmed} \\
	Department of Mechanical Engineering\\
	Massachusetts Institute of Technology\\
	Cambridge, MA 02139 \\
	\texttt{faez@mit.edu}
}

\renewcommand{\shorttitle}{C-ShipGen}

\hypersetup{
pdftitle={C-ShipGen-DDPM},
pdfsubject={},
pdfauthor={Noah J.~Bagazinski, Faez~Ahmed},
pdfkeywords={Naval Architecture; Ship Hull Design; Generative Artificial Intelligence; Conditional Denoising Diffusion Probabilistic Model; Drag Reduction; Parametric Design}
}
\begin{document}

\maketitle

\begin{abstract}
 Ship design is a complex design process that may take a team of naval architects many years to complete. Improving the ship design process can lead to significant cost savings, while still delivering high-quality designs to customers. A new technology for ship hull design is diffusion models, a type of generative artificial intelligence. Prior work with diffusion models for ship hull design created high-quality ship hulls with reduced drag and larger displaced volumes. However, the work could not generate hulls that meet specific design constraints. This paper proposes a conditional diffusion model that generates hull designs given specific constraints, such as the desired principal dimensions of the hull. In addition, this diffusion model leverages the gradients from a total resistance regression model to create low-resistance designs. Five design test cases compared the diffusion model to a design optimization algorithm to create hull designs with low resistance. In all five test cases, the diffusion model was shown to create diverse designs with a total resistance less than the optimized hull, having resistance reductions over 25\%. The diffusion model also generated these designs without retraining. This work can significantly reduce the design cycle time of ships by creating high-quality hulls that meet user requirements with a data-driven approach.
\end{abstract}

\section{Introduction}
Generative artificial intelligence has shown to be a promising tool for engineering design. By training models on engineering datasets, generative models have created designs with high performance. These tools are particularly useful in ship design, as the complexity of balancing competing trade-offs in a ship requires long design cycles for human design teams. A hull's shape affects several key aspects of a ship's performance, including buoyancy, upright stability, hydrodynamics, and general arrangements. A generative model specifically trained to generate ship hulls can improve this workflow by creating high-quality designs quickly and inexpensively. The availability of open-source datasets on ship hull designs enables the use of generative artificial intelligence for design. This work builds on prior work, a guided diffusion model called \textit{ShipGen}~\cite{bagazinski2023shipD}.

Hull design was chosen as the application for this work as hulls have a direct impact on over 70\% of the cost of a ship~\cite{lin2017feature} and they are the first step in the traditional workflow for ship design~\cite{evans1959basic}. Ship hulls can exist across many length scales, ranging from a few meters to over three hundred fifty meters in length. In addition, hulls can exhibit large ranges in relative dimensions such as the beam, draft, depth, and volume displacement. A well-designed generative model for hull design should consider the scale and diversity of ship hulls in its training. This would allow the generative model to create ship hulls based on a designer's needs.  

\begin{figure}[H]
\begin{center}
\setlength{\unitlength}{0.012500in}%
\includegraphics[width = 6.5in]{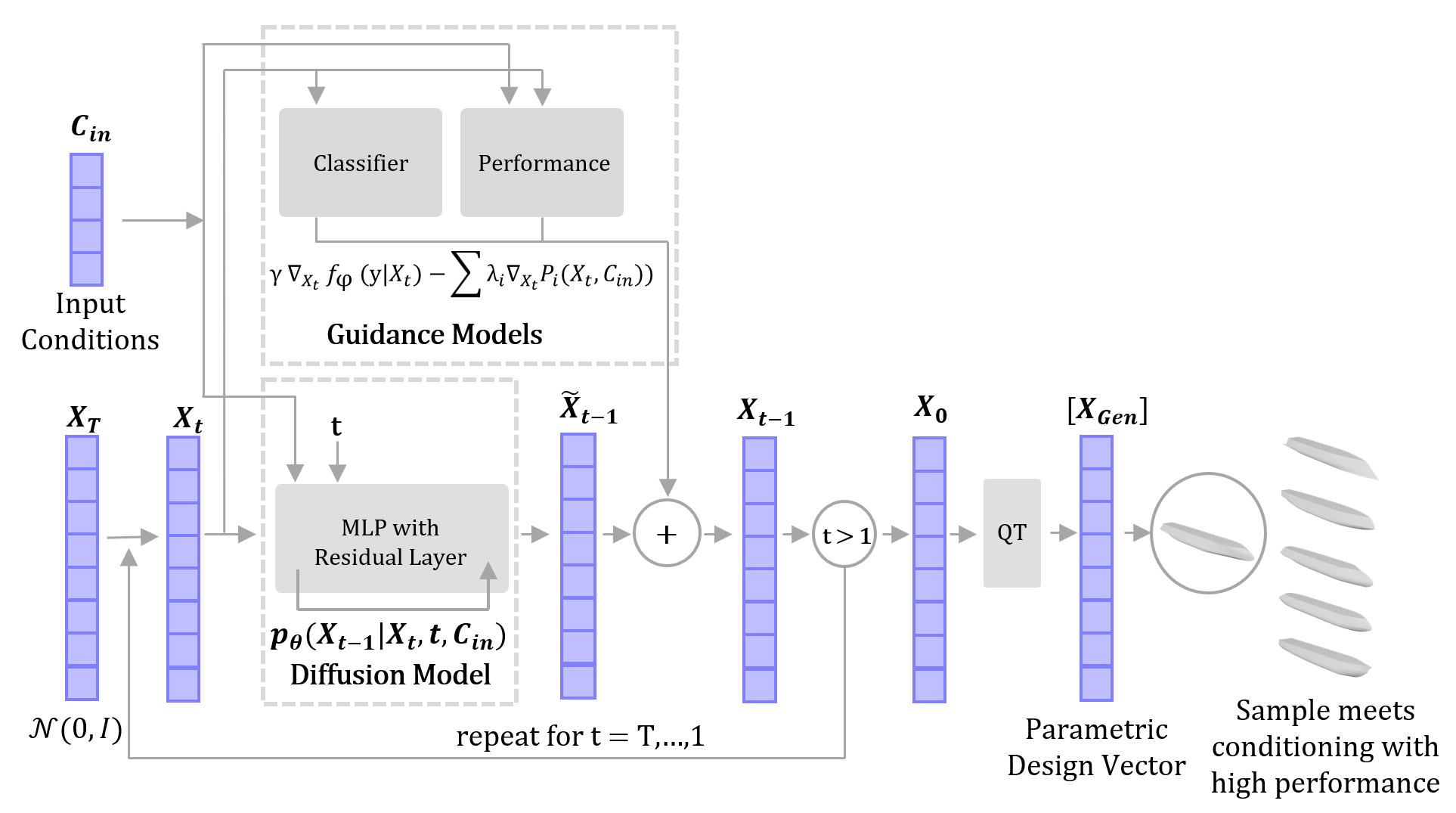}
\caption{C-ShipGen is a guided conditional diffusion model that generates hull designs with low resistance while maintaining the principal dimensions provided by the user during sampling. The model leverages guidance gradients from pre-trained regression models to improve the performance of the hulls.}
\label{fig:figure_Overview} 
\end{center}
\end{figure}

This work proposes a model called C-ShipGen, which generates early-stage hull designs considering a designer's inputs: length, beam, draft, depth, volume displacement, and intended velocity. C-ShipGen is a conditional diffusion model that implements guidance algorithms to create ship hull designs with low resistance while constraining to a user's desired principal characteristics.  Figure~\ref{fig:figure_Overview} shows an overview of C-ShipGen, highlighting how the model utilizes a combination of input conditioning and guidance algorithms during the design sampling process. The following sections detail prior research on generative artificial intelligence for engineering design and computational ship design; the methods for training and sampling hull designs with C-ShipGen; the evaluation of hulls generated by C-ShipGen; and a discussion on the work. C-ShipGen generates ship hulls with low resistance for future design analysis. These generated hulls do not necessarily resemble real-world ship hulls as many other factors influence the design of hulls in addition to total resistance. Through the development of C-ShipGen, the contributions of this paper are:

\begin{enumerate}
    \item The use of a conditional diffusion model to generate diverse ship hulls within a 5\% volume error tolerance given desired principal characteristics across the full spectrum of hull sizes and relative dimensions found in real-world hull designs. 
    \item The use of guidance in a conditional diffusion model to generate hulls with lower total resistance than optimized hulls, having resistance reductions greater than 25\% while maintaining the displaced volume within 5\% of a target. 
    
\end{enumerate}

\section{Prior Work}    
Computational ship design refers to the application of computer-based modeling, simulation, and optimization techniques in the design and analysis of marine vessels. This facilitates more efficient, innovative, and integrated design solutions.
Historically, computational ship design can be divided into three categories: design representation, forward modeling which includes surrogate regression models, and inverse design or synthesis, which includes optimization methods. Recently, generative artificial intelligence methods have emerged as a powerful technique, which has been used to represent and synthesize ship hull designs. 

In order to computationally design a product, a design representation is needed that allows a computer to manipulate the design. For ship design, the two most popular modes to represent the design of a ship hull are parameterized vectors~\cite{brown2003multiple, HullOpt, read2009drag,  zhang2018parametric, chrismianto2014parametric, lu2016hydrodynamic, PSOShip_Opt, PSO_Multi_Opt, hodgesai, bagazinski2023shipD} and free form deformation techniques~\cite{wang2022shipEncoding,ao2021artificial,ao2022artificial,peri2001design,demo2021hull,abbas2023deepmorpher}. With a design representation, a dataset of designs can can be created by calculating or simulating performance metrics for each design.  With the dataset, data-driven models can be trained to make inferences on new designs. For ship design specifically, the works of Khan et al.~\cite{khan2022shape, khan2022geometric,khan2023shiphullgan},  Shaeffer et al.~\cite{shaeffer2023application, shaeffer2020application}, and Bagazinski et al.~\cite{bagazinski2023shipD} have looked at various methods to create diverse design spaces and datasets for ship hull design. 

With a dataset of designs and performance metrics of a hull as inputs, data-driven surrogate models, using methods such as neural networks, provide a computationally inexpensive way to predict performance. Significant research has been done on predicting the hydrodynamics of hulls with neural network-based surrogates due to the high cost of performing computational fluid dynamics simulations~\cite{ao2021artificial, ao2022artificial, khan2022geometric, khan2022shape, peri2001design, read2009drag, lu2016hydrodynamic, HullOpt,wang2022shipEncoding, marlantes2021Modeling, silva2023implementation}.

Generative artificial intelligence models are designed to create new content or data that resemble the input they were trained on. These models can produce a wide range of outputs, from text to images, and even complex design structures. In the context of ship design, datasets facilitate quick performance predictions and serve as a foundation for training generative artificial intelligence models to innovate in hull design. Prior work has explored various generative approaches, including 
variational autoencoders~\cite{hodgesai}, generative adversarial networks~\cite{yonekura2023designing, khan2023shiphullgan}, and diffusion models~\cite{bagazinski2023shipgen}, each contributing uniquely to the field. 

Diffusion models, in particular, iteratively modify a noisy data vector over many specified steps. This transforms random data to mirror the statistics of training data~\cite{ho2020denoising}. The development of diffusion models has shown that they can generate complex data and already have applications for engineering design. For example, diffusion models were shown to create higher quality images compared to generative adversarial networks~\cite{ho2020denoising}. Subsequent advancements in diffusion models introduced guidance, where gradients from a classifier neural network guide image synthesis to match a specific image classification label~\cite{dhariwal2021diffusion}. This evolution enabled text-to-image diffusion models that employ text-based guidance to craft custom, lifelike images~\cite{ramesh2022hierarchical,rombach2021highresolution}. Guided diffusion models have found applications in generating 3D shapes from image data~\cite{liu2023zero1to3}. In addition, guided diffusion can be applied to engineering design generation. For example, guided diffusion has been used to create two-dimensional structures~\cite{maze2022topodiff,giannone2023aligning,giannone2023learning}, room layouts~\cite{ploennigs2023diffusion}, thermometers~\cite{yang2023product}, architected materials~\cite{lew2023single},  and vehicles~\cite{arechiga2023drag}. 
A notable advantage of diffusion models is their adaptability, allowing the incorporation of new design constraints or objectives without necessitating retraining the entire model. This attribute is particularly beneficial for iterative design processes, where continuous adjustments are essential for optimizing performance. 

The work presented in this paper builds on the prior work in training diffusion models for hull design. A prior model called \textit{ShipGen} can generate parametric ship hull designs that have 91.4\% lower wave drag and 47.9x higher internal volume on average compared to the original training data~\cite{bagazinski2023shipgen}. \textit{ShipGen} implements seven different guidance models to create these designs.  A major shortcoming of this model is that there is no control to generate desired principal characteristics in a hull design, such as length, beam, draft, depth, and displacement. All hull designs generated by \textit{ShipGen} are purely influenced by the guidance models and not by a human designer. This makes \textit{ShipGen} impractical for some real-world design applications, where designers may seek more freedom in imposing constraints. To address this gap, we propose a model that utilizes an additional feature called conditioning, where the diffusion model is given a hull design's principal characteristics during training. This way, the diffusion model will generate designs that satisfy the principal characteristics provided by a designer during sampling. The following subsections detail this improvement and showcase a few design applications with this diffusion model. 

\section{Methods}
This section outlines the methodology for developing C-ShipGen (Conditional \textit{ShipGen}). The first subsection explores the training dataset of ship hulls, including formulae for geometric and performance measurements of the dataset hulls. The second subsection details the model training for the regression models used in this study. The third subsection details training and sampling with a conditional diffusion model. The fourth subsection details the methodology for analyzing designs generated with the diffusion model. 

\subsection{Dataset}
To train the regression and generative models, a dataset of ship hulls and their respective performance metrics was created. The training dataset consists of 82,168 parametric ship hull designs. These dataset hulls were derived from the following sources: 
\begin{itemize}
    \item 30,000 hulls from the \textit{Ship-D} dataset~\cite{bagazinski2023shipD}
    \item 41,752 hulls generated using \textit{ShipGen}~\cite{bagazinski2023shipgen}
    \item 10,416 hulls subset from the former \textit{ShipGen} hulls with the addition of randomly parameterized bulbous bows and sterns 
\end{itemize}

\begin{figure}[ht]%
\centering
\begin{minipage}[t]{.48\linewidth}
    \centering
    \includegraphics[width = 3in]{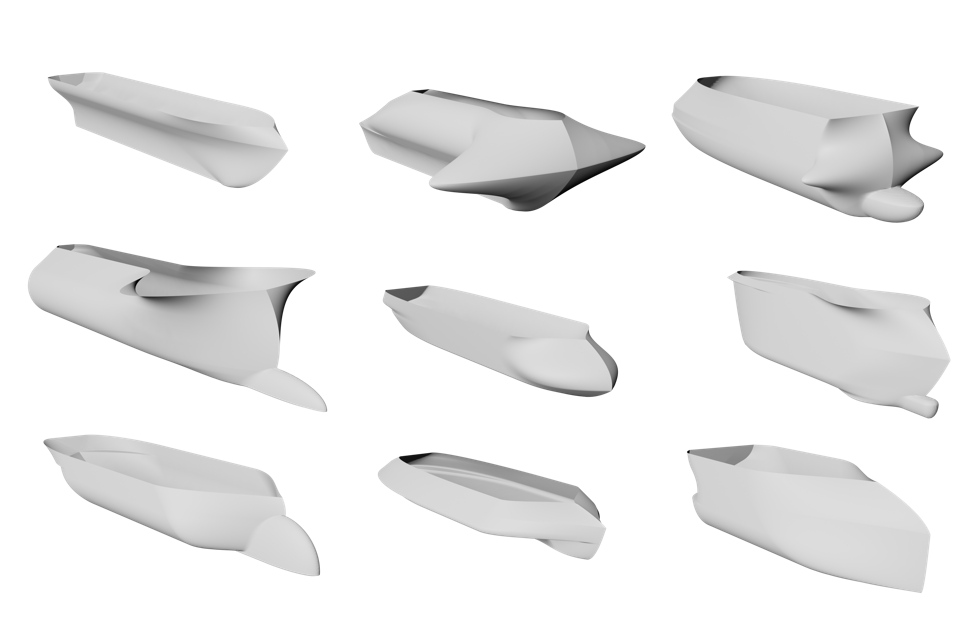}
    \captionof{figure}{A selection of hulls from the \textit{Ship-D} dataset, showing the variability possible with the hull parameterization. A random sampling from the dataset may lead to unrealistic hulls, containing combinations of features that do not resemble real-world ships and features that lead to poor performance.}
    \label{fig:figure_Ship-DHulls} 
\end{minipage}\hfill%
\begin{minipage}[t]{.48\linewidth}
    \centering
    \includegraphics[width=3.25in]{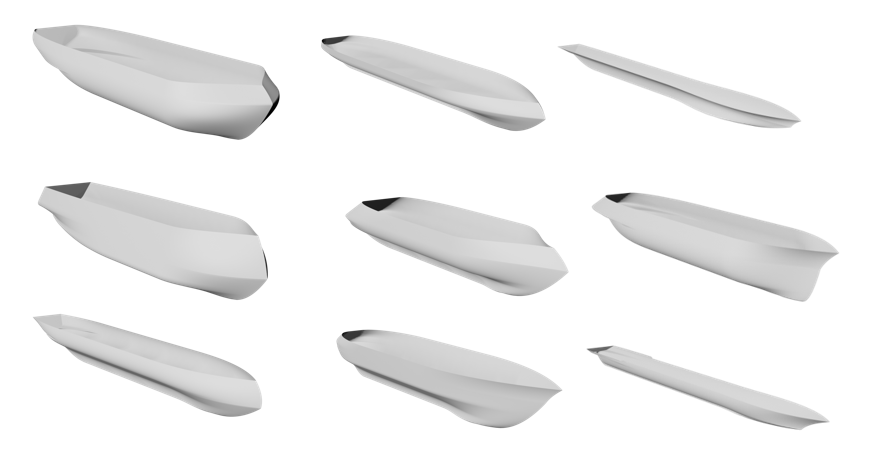}
    \captionof{figure}{A selection of hulls generated with multi-objective guided performance generation. Notice the relative slenderness of the hulls leading to drastically reduced drag coefficients relative to the \textit{Ship-D} dataset hulls.}
    \label{fig:figure_GenHulls} 
\end{minipage}
\end{figure}

These parametric hull designs are represented with a forty-five-parameter scheme that algebraically defines the hull's surface. The \textit{Ship-D} hulls cover the full design space possible with the parametric design scheme. The \textit{Ship-D} hulls do not represent realistic-looking or performing hullforms. The hullforms, however, provide a large diversity of feature combinations that encompass realistic hull designs for machine learning applications~\cite{bagazinski2023shipD}. A few examples of the \textit{Ship-D} dataset hulls are shown in Figure~\ref{fig:figure_Ship-DHulls}. Using the initial 30,000 \textit{Ship-D} hulls, a guided tabular diffusion model called \textit{ShipGen} was trained to generate high-performing hull designs. The mean performance of the 41,752 \textit{ShipGen} hulls and the mean performance of the 30,000 \textit{Ship-D} hulls was calculated and non-dimensionalized. Comparatively, the generated \textit{ShipGen} hull designs have a mean wave drag that is 91.4\% lower and a mean internal volume that is 47.9x higher than the mean performances found among the  \textit{Ship-D} dataset hull designs~\cite{bagazinski2023shipgen}. A selection of \textit{ShipGen} hulls is shown in Figure~\ref{fig:figure_GenHulls}. The \textit{ShipGen} hulls are much more representative of realistic hull designs compared to the \textit{Ship-D} hulls. Among the \textit{ShipGen} hulls, it was observed that few were generated with bulbous bows and bulbous sterns, a feature that can reduce the drag on hulls when designed well. To increase the presence of bulbs in the training dataset, bulbs were added to 10,416 hulls by randomly sampling the design parameters for bulbs. These bulbs were not tuned for hydrodynamic performance. Only a smaller subset of the \textit{ShipGen} hulls were selected for bulbs as these were the hull designs that allowed for the generation of feasible bulb designs. 

The feasibility of hulls is calculated using a set of forty-nine algebraic constraints. These constraints ensure that a parametric hull surface is watertight and non-self-intersecting. These algebraic constraints are solved with the parameter values for a given hull design without generating the hull's surface, reducing the total computation time. Full documentation of the hull design parameters and constraints is provided at ~\url{https://decode.mit.edu/projects/ShipGen/}.

\subsubsection{Measures of Hull Geometry}
In addition to the parametric hull designs, the dataset includes geometric measures for each hull. The displaced volume, wetted surface area, and waterline length were calculated at 100 evenly spaced draft marks across the depth of each hull. To generate designs regardless of scale, these geometric measures are scaled using the first parameter in the hull representation: length overall, or $LOA$. The equations for the normalized volume, surface area, and waterline length are provided in Equations~\ref{eq_NormVt},~\ref{eq_NormSAt}, and~\ref{eq_NormWLt}.  
\begin{equation}
 V_{t*} =  \left( \frac{\int_{0}^{T/D = t*} \delta V(z) \, \delta z} {\text{LOA}^3} \right)
\label{eq_NormVt}
\end{equation}
\begin{equation}
 SA_{t*} = \left( \frac{\int_{0}^{T/D = t*} \delta SA(z) \, \delta z} {LOA^2} \right)
\label{eq_NormSAt}
\end{equation}
\begin{equation}
 WL_{t*} = \left( \frac{X_{fwd}(t*) - X_{aft}(t*)} {LOA} \right)
\label{eq_NormWLt}
\end{equation}
In each equation, $LOA$ scales the value based on its dimensionality: $LOA^n$. The other terms, $t*$ is ratio of draft, $T$, to depth, $D$. During model training, $t*$ will be used as an additional embedding to hull shape to predict the geometric measures of a hull at a specific draft mark.  

\subsubsection{Calculation of Total Hull Resistance}
In addition to the geometric measures of each hull, the total resistance of each hull is calculated for many speed and draft conditions. Total resistance, $R_T$ is estimated to be the sum of wave-making resistance, $R_w$, and skin friction resistance, $R_f$, as seen in Equation~\ref{eq_Rt}. Wave drag calculations were computed using Michell's integral. Michell's integral was chosen as the simulation for this study as it considers the full 3D geometry of a hull in the total resistance prediction while being computationally inexpensive to calculate. In practice, any fluid simulation would work for this study as the methodology of training generative models is the same. For this study, 2.6 million fluid simulations were performed using the Michell integral. The Michell integral balances the need for accurate simulation data with reduced computational cost of performing the simulation. For this study, thirty-two wave drag coefficients for each hull across four different drafts and eight velocity conditions. The four drafts are $t*$ = 0.25, 0.33, 0.50, and 0.67. The eight velocity conditions are normalized using Froude scaling as seen in Equation~\ref{eq_Fn}, where $g$ is gravitational acceleration, $U$ is the ship's speed, and $WL_{t*}$ is the non-dimensional waterline length for a given draft. In the denominator, $WL_{t*}$ is multiplied by $LOA$ to balance the dimensions of $U$ and the denominator.
 \begin{equation}
R_T = R_w + R_f
\label{eq_Rt}
\end{equation}
\begin{equation}
F_n = \frac{U}{\sqrt{g WL_{t*} LOA}}
\label{eq_Fn}
\end{equation}
The eight ship speeds are scaled between $F_n =0.10$ and $F_n = 0.45$ in increments of 0.05, corresponding to typical operating conditions of traditional displacement hulls~\cite{AppNavArch,newman2018marine}. Given the 3D surface of a hull submerged to $t*$ and a velocity, $U$, the wave-making resistance is calculated with Equation~\ref{eq_Mich}.  
\begin{equation}
 R_w = \frac{A \rho g^2}{\pi U^2} \int_{1}^{\infty} (I^2 + J^2) \frac{\lambda^2}{\sqrt{\lambda^2 - 1}} \, d\lambda
\label{eq_Mich}
\end{equation}
 where $\rho$ is the density of water, and $A$, $I$, and $J$ are integrated terms relating to the surface normal across the hull and the direction of wave propagation. Further insight into these terms is in Michell's paper from 1898~\cite{michell1898Wave}. With these thirty-two wave drag measurements, a given hull's wave-making resistance for a given $t*$ and $F_n$ is interpolated between these calculations. 

 Skin friction resistance is calculated using the ITTC-1957 formula in Equations~\ref{eq_Cf} and~\ref{eq_Rf}. 

 \begin{equation}
C_f = \frac{0.075}{(\log(Re) - 2)^2}
\label{eq_Cf}
\end{equation} 
\begin{equation}
R_f = \frac{1}{2}C_f\rho U^2 SA_{t*} LOA^2
\label{eq_Rf}
\end{equation}
The Reynolds number of the hull, $Re$, scales with forward velocity, $U$, and waterline length, $WL_{t*}$.  $SA_{t*}$ is the non-dimensionalized wetted surface area of the hull for a given draft. Together,  $R_w$ and $R_f$ can be used to calculate the coefficient of total resistance, $C_T$. To learn with the dataset, $C_T$ is scaled by $LOA^2$ as opposed to the more traditional use of wetted surface area. This was done so that a regression model can embed $LOA$ and $t*$ to predict $C_T$ without explicitly providing $SA_{t*}$. Additionally, $C_T$ is on a logarithmic scale so that the distribution of $C_T$ is approximately Gaussian for model training. The calculation of $C_T$ is provided in Equation~\ref{eq_CT}.

 \begin{equation}
  C_{T} = \log_{10} \left( \frac{R_w + R_f}{\frac{1}{2} \rho U^2 LOA^2} \right)
\label{eq_CT}
\end{equation}  

With this representation, $C_T$ will be predicted using the 45 design parameters, $t*$ for a draft embedding, and $F_n$ for a speed embedding. The predicted coefficient of total resistance, $\hat{C_T}$, can then be scaled to a prediction of total resistance. $\hat{R_T}$ with Equation~\ref{eq_RT_pred}.

 \begin{equation}
  \hat{R_{T}} = 10^{\hat{C_T}} \frac{1}{2} \rho U^2 LOA^2
\label{eq_RT_pred}
\end{equation}  

The following subsection will detail training the regression models and the diffusion model with a dataset made from these equations. 

\subsection{Regression Modeling with Neural Networks}
Using neural networks, four regression models were trained with the dataset: displaced volume, coefficient of total resistance, waterline length, and design feasibility. A trained neural network for regression provides two key benefits for computational design. The first benefit is a fast prediction of performance directly from design parameters. The second benefit is the ability to calculate the gradient of a performance metric with respect to the design parameters. This subsection will detail the process of training the regression models. 

Following prior work, the forty-four (not including $LOA$), design parameters are quantile normalized and then scaled between -1 and 1~\cite{bagazinski2023shipgen}. Quantile normalization bins values of the design parameters so that the distributions of the design parameters are approximately Gaussian. This parameter scaling improves the diffusion model training. Scaling the parameters for the regression models allows them to work in conjunction with the diffusion model during sampling. This process, called guidance, will be described in the following subsection. 

The process for training the coefficient of the total resistance regression model is described in Table~\ref{table:trainingCT}. In the algorithm, the neural network is represented as $P_{C_T}$. The inputs to the regression model are a quantile normalized design vector, $X_i$, a draft embedding, $t*$, a speed embedding, $F_n$, and a length embedding, $\log(LOA)$. The loss function is the mean-squared error loss between the ground truth and the prediction of $C_T$. The draft and speed embeddings are restricted to the limits of the range of draft and Froude numbers used in the dataset calculation of wave-making resistance. The length embedding is on a logarithmic scale for hulls with a length between 3 meters and 450 meters. This range of values allows the model to learn the relative influence of skin friction across velocity and length scales. Given the diversity of hull forms in the dataset and the range of length scales, this regression model is trained to predict the coefficient of total resistance on a large diversity of shapes, speeds, and sizes. This model was trained with a batch size of 1024 for 50,000 batches. The resistance prediction model predicts the total resistance coefficient across the full spectrum of the dataset hulls, draft ratios, and Froude numbers. This regression model predicts the total resistance coefficient derived from the simulation with an $R^2$ of 0.997. Results of the training accuracy relative to the simulation data are shown in Figure~\ref{fig:figure_Study0_RT} and Figure~\ref{fig:figure_testcases_RT}. 

\begin{table} [ht]
\begin{center}
\begin{tabular}{l l} 
\hline
1: & \textbf{repeat}   \\            
2: & \textbf{select} $X_i$ from Dataset \\ 
3: & $t* \sim Uniform([0.25,0.67])$ \\
4: & $F_n \sim Uniform([0.05,0.45])$ \\
5: & $\log(LOA) \sim Uniform([0.47, 2.65])$ \\
6: & \textbf{interpolate} $R_{w_{t*,F_n}}$, $SA_{t*}$, and $WL_{t*}$ from Dataset \\ 
7: & \textbf{calculate} $C_f$, $R_f$, and $C_T$ \\ 
8: & $\hat{C_{T}} = P_{C_T}\left( X_i, t*, F_n, \log(LOA) \right)$ \\
9: & Take gradient descent step on: \\
 & $\nabla_{P_{C_T}}(C_T -\hat{C_{T}})^2$ \\
10: & \textbf{until} converged\\
\hline
\end{tabular}
\end{center}
\caption{The training algorithm for the total resistance coefficient regression model. The algorithm randomly samples a Froude number and draft ratio for a hull in each batch so the model is trained on a full spectrum of speeds and drafts for the hulls in the dataset.}
\label{table:trainingCT}
\end{table}

The training for the volume regression model, $P_{V}$, and the waterline regression model, $P_{WL}$, is similar. The volume regression model was trained to predict $\log(V_{t*})$ given $X_i$ and $t*$. The model was trained to predict the logarithm of volume as this term has an approximate Gaussian distribution across the dataset. Similar to the $C_T$ regression model, the volume regression is trained to predict the displaced volume for any draft on a large diversity of hullforms. The waterline regression model predicts the waterline length of a given hull at a given draft. This model is used with the coefficient of total resistance regression model when $WL_t*$ is unknown. 

The final regression model is a feasibility classifier, $f_{\phi}$. The feasibility classifier provides guidance gradients to influence the diffusion model to generate parametric designs that satisfy the forty-nine algebraic feasibility constraints~\cite{bagazinski2023shipgen}. The only input to the feasibility classifier is a design vector, $X_i$. To learn the distinction between feasible and infeasible designs, a set of 82,793 design vectors that violate at least one constraint was generated. The loss function in training was binary cross-entropy loss, which is better for classifier training than mean-squared-error loss.

\subsection{Conditional Diffusion Model}
\begin{figure}[ht]
\begin{center}
\setlength{\unitlength}{0.012500in}%
\includegraphics[width = 6.5in]{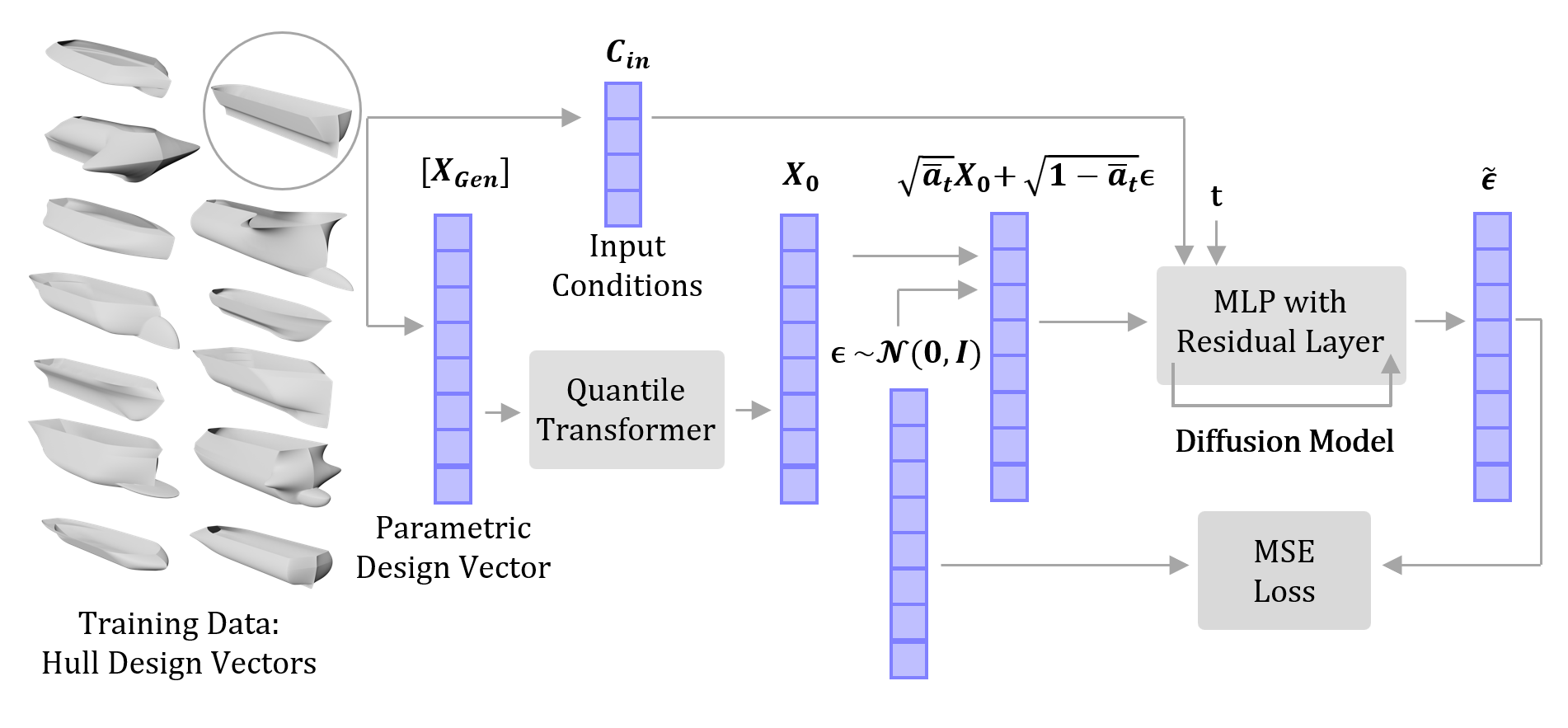}
\caption{During training, the diffusion model predicts a denoising step, given a timestep embedding and a partially noised sample design vector. The model is informed by the input conditioning at each denoising step.}
\label{fig:figure_DDPM_Training} 
\end{center}
\end{figure}

A diffusion model is a generative artificial intelligence model that generates new instances of data by denoising random information over many steps. The generated sample will fall within the statistical distribution of the training dataset samples. Conditional diffusion models are similar to the standard diffusion model described by ~\cite{ho2020denoising}, however, their structure includes extra layers that embed information in the training and sampling process. This conditional diffusion model is a modified version of the \textit{ShipGen} model, a tabular diffusion model for ship hull design~\cite{bagazinski2023shipgen}. The conditioning for the model is the principal characteristics of the hull: draft, beam, depth, and displaced volume. The diffusion model and the conditioning use parameters that are scaled by the length overall, which is the first term in the forty-five parameter representation used to generate the hull designs. After sample generation, the parameter terms are re-scaled by $LOA$ to measure the scaled design. By training with parameters scaled with respect to $LOA$, the model does not have to learn ``length" in addition to the statistical relationships between the parameters to generate a hull design. This reduces the complexity of the learning task. The training algorithm for the conditional diffusion model is stated in Table~\ref{table:DDPMTrainAlg} and illustrated with Figure~\ref{fig:figure_DDPM_Training}.

\begin{table} [ht]
\begin{center}
\begin{tabular}{l l} 
\hline
1: & \textbf{repeat} \\
2: & $X_0 \sim q(X_0)$ \\
3: & $t* \sim Uniform([0.01,1.0])$ \\
4: & \textbf{interpolate} $V_{t*}$, $B$, and $D$ from Dataset \\
5: & $C = [t*,\log(V_{t*}),B,D]$ \\
6: & $t \sim Uniform(\{1,...,T\})$ \\
7: & $\epsilon \sim N(0,I)$ \\
8: & Take gradient descent step on: \\
 & $\nabla_{\theta}||\epsilon - \epsilon_{\theta}(\sqrt{\bar{a_t}}X_0 + \sqrt{1 - \bar{a_t}}\epsilon, t, C)||^2$ \\
9: & \textbf{until} converged\\
\hline
\end{tabular}
\end{center}
\caption{This is the training algorithm for a conditional diffusion model. The diffusion model is represented by the function $\epsilon_{\theta}(X_0,\epsilon,t, C)$ in step 8. The conditioning for hull design is the draft,$t*$; displaced volume, $\log(V_{t*})$; beam, B; and depth, D, of the hull, $X_0$}
\label{table:DDPMTrainAlg}
\end{table}

The diffusion model is conditioned with the draft, $t*$; volume, $V_{t*}$; Beam, $B$, and depth, $D$ of each hull. During training, the original design vector is partially noised to a timestep, $t$, and the conditional diffusion model predicts the noise of the sample at that timestep. Conditioning is applied to the diffusion model to influence the denoising process to satisfy input conditioning. To clarify in the algorithm, $t$ is the timestep embedding of the denoising process, while $t*$ is the draft embedding of the hull design in the conditioning vector. 

After training, the diffusion model can be sampled to create design vectors that satisfy the input conditioning. While the diffusion model can generate hull designs that satisfy the input conditioning, the sampling process does not consider the total resistance of the hull. To generate hulls with reduced total resistance, the total resistance coefficient regression model is implemented in the sampling process as a guidance algorithm. Guidance leverages the gradients of the regression model at each timestep to influence the denoising process toward producing designs with reduced total resistance. In addition to resistance guidance, the feasibility classifier and the volume prediction regression models are also used as guidance. The feasibility classifier aims to improve the likelihood that a generated design vector leads to a feasible hull design. The volume guidance assists the diffusion model in generating a design that satisfies the input conditioning for displaced volume. This was implemented to prevent the resistance guidance from over-influencing the sampling process and producing hulls that do not satisfy the input conditioning. Each guidance algorithm is tuned with a hyperparameter: $\gamma$ tunes the classifier guidance, while $\lambda_0$  and $\lambda_1$ tune the performance guidance. The sampling process is illustrated in Figure~\ref{fig:figure_Overview}. The sampling algorithm is stated in Table~\ref{table:DDPMSampAlg}.

\begin{table} [ht]
\begin{center}
\begin{tabular}{l l} 
\hline
1: & \textbf{input} $C = [t*,V,B,D,]$ and $U, LOA$ \\ 
3: & $X_T \sim N(0,I)$\\
4: & \textbf{for} $t = T,...,1$ \textbf{do} \\
5: & $ Z \sim N(0,I)$ if $t > 1$, else $z = 0 $ \\
6: & $F_n = \frac{U} {g P_{WL}(X_t,t*)LOA} $\\
7: & $  X_{t-1} = \frac{1}{\sqrt{\alpha_t}} \left(X_t - \frac{1 - \alpha_t}{\sqrt{1 - \bar{\alpha}_t}} \epsilon_{\theta}(X_t ,t, C)\right) + \sigma_t(Z(1-\gamma)) + \gamma \nabla_{X_t} f_{\phi}(y|X_t)$  \\
&\multicolumn{1}{r}{$ - \lambda_0\nabla_{X_t}P_{C_T}\left( X_i, t*, F_n, \log(LOA) \right) - \lambda_1\nabla_{X_t}\left(V - P_{V}(X_t,t*)\right)^2 $}\\
8: & \textbf{end for} \\
9: & \textbf{return} $X_0$ \\
\hline
\end{tabular}
\end{center}
\caption{This is the sampling algorithm for a guided conditional diffusion model. The diffusion model is represented by the function $\epsilon_{\theta}(X_t,t,C)$ in step 7. }
\label{table:DDPMSampAlg}
\end{table}

In the sampling algorithm, $\gamma$ is equal to 0.2, while $\lambda_0$ and $\lambda_1$ are equal to 0.3. The performance guidance hyperparameters are set equal so that no model overpowers the others. The $\lambda$ values are set low so the guidance models do not overpower the denoising process from the diffusion model. One advantage to leveraging performance guidance is that the diffusion model does not need to be retrained to produce designs when considering different objectives. The guidance model can simply be replaced with a new one. Using an \textit{NVIDIA GeForce RTX 4090}, 512 samples are generated in approximately 2.5 seconds. The feasibility check, total resistance calculation, and geometric measurements are computed on a single \textit{Intel Core i9-13900K} core in approximately 2.5 seconds per sample. After sampling, the parallel CPU process across 32 cores for the 512 samples is less than 30 seconds. 

Diffusion models rely on a degree of randomness in the denoising process. Sampling from the guided conditional model will not guarantee that every generated design will be high-quality. Therefore, studies on designs generated from the model evaluate the statistics from a set of generated designs. In addition, high-quality designs will be filtered from a set of generated designs to select potential candidates for further design evaluation. This is distinctly different from other data-driven design approaches, such as optimization, which provides some guarantee of design performance and constraint satisfaction among generated designs. Assuming that intended designs fall within the statistical distribution of the training dataset, conditional diffusion models can produce a large diversity of designs without needing to retrain the model, significantly decreasing the computational effort to create high-quality designs compared to optimization methods. 

\subsection{Evaluation of Diffusion Model for Low-Drag Design}
A baseline comparison is needed to evaluate the ability of the guided conditional model to generate high-quality hull designs. Design optimization for drag reduction was selected as the baseline comparison. To conduct the study, five design test cases compare optimized hull forms to diffusion-generated hull forms. The purpose of this test is twofold:

\begin{enumerate}
    \item Evaluate the diffusion model's ability to design low-resistance hulls while meeting specific dimensional properties.
    \item Evaluate the accuracy of the total resistance regression model for a wide array of designs, scales, and relative speeds. 
\end{enumerate}

The five test cases were selected to create a unique set of dimensional requirements for both the design optimization and the diffusion model to satisfy. The designs of real-world ship classes inspired the dimensions of the five test cases. The design inspirations are a supercarrier~\footnotemark[1] , a kayak~\footnotemark[2] , a NeoPanamax container ship~\footnotemark[3] , a frigate~\footnotemark[4] , and a ROPAX ferry~\footnotemark[5] . These test cases encompass two military-style ships, two large ship designs, one small hull design, two small block coefficient designs, and one high beam-to-draft ratioed hull. The principal dimensions and design speed of the test cases are provided in Table~\ref{table:HullTestcases}. 

\begin{table}[ht]
\begin{center}
\begin{tabular}{l r r r r r r r r} 
\hline
\multicolumn{1}{l}{\multirow{2}{*}{\textbf{Test Case}}} & \multicolumn{1}{c}{\textbf{LOA}} & \multicolumn{1}{c}{\textbf{BOA}} & \multicolumn{1}{c}{\textbf{T}} & \multicolumn{1}{c}{\textbf{D}} & \multicolumn{1}{c}{\boldmath$\volume$} & \multicolumn{1}{c}{\boldmath$C_B$} & \multicolumn{2}{c}{\boldmath$U_s$} \\
& \multicolumn{1}{c}{$(m)$} & \multicolumn{1}{c}{$(m)$} & \multicolumn{1}{c}{$(m)$} & \multicolumn{1}{c}{$(m)$} & \multicolumn{1}{c}{$(m^3)$} & \multicolumn{1}{c}{$(-)$} & \multicolumn{1}{c}{$(m/s)$} & \multicolumn{1}{c}{$(knots)$} \\
\hline
Supercarrier~\footnotemark[1] &     333.0 &     42.1 &  11.3 &  29.6 &  97,561 &    0.617 &     16.0 &  31.1\\
Kayak~\footnotemark[2] &             3.8&        0.787&  0.15&   0.438&  0.166  &    0.372 &     1.50 & 2.92\\
NeoPanamax~\footnotemark[3] &        366.0  &    50.0 &  15.2 &  40.0&   182,114 &   0.654 &     10.3 & 20.0\\
Frigate~\footnotemark[4] &           127.0   &   16.0&   6.90&   11.0&   4,488 &     0.320 &     14.4 & 28.0 \\
ROPAX Ferry~\footnotemark[5] &       72.0 &      20.0 & 3.2 &    4.8 &   3,917  &    0.850&      6.17 &  12.0\\
\hline
\end{tabular}
\end{center}
\caption{This table provides the dimensions of hull design test cases inspired by real-world ship designs. These test cases cover a diversity of principal characteristics, hull speeds, and length scales.}
\label{table:HullTestcases}
\end{table}

\footnotetext[1]{supercarrier Inspiration: \url{https://www.nvr.navy.mil/SHIPDETAILS/SHIPSDETAIL_CVN_68.HTML}}
\footnotetext[2]{Kayak Inspiration: \url{https://oldtownwatercraft.johnsonoutdoors.com/us/shop/kayaks/recreation/loon-126}}
\footnotetext[3]{NeoPanamax Inspiration: \url{https://www.cmacgm-group.com/en/group/at-a-glance/fleet/ships/9780873/cma-cgm-t-roosevelt}}
\footnotetext[4]{Frigate Inspiration: \url{https://www.dcms.uscg.mil/Our-Organization/Assistant-Commandant-for-Acquisitions-CG-9/Programs/Surface-Programs/National-Security-Cutter/}}
\footnotetext[5]{ROPAX Inspiration: \url{https://www.steamshipauthority.com/about/vessels}}

It is not expected for the diffusion model nor the optimization algorithm to produce designs that look like real-world ship designs with the same principal characteristics. Real-world hull designs satisfy many additional performance objectives in addition to total resistance; such as seakeeping, upright stability, cargo packing, general arrangements, etc. Since these hulls are generated only considering total resistance and principal dimensions, they are not expected to resemble real-world hull designs. For each design test case, 512 hull designs were generated with the full diffusion model, 512 were generated with the diffusion model and classifier guidance only, and 100 designs were generated using design optimization. In each test case, the diffusion-generated designs will be evaluated on drag, dimensional target satisfaction, and design diversity compared to optimized designs.

The optimization algorithm used for these studies is \textit{NSGA-II}~\cite{Deb2002NSGA2}. \textit{NSGA-II} is a state-of-the-art genetic algorithm for optimizing two or more objectives. Genetic algorithms are a set of optimization algorithms that act similarly to biological evolution to drive the optimization over several ``generations". These tests were performed with a population of 100 samples for 200 generations. The initial population consisted of randomly selected designs from the dataset. For each test case, the optimization algorithm constrains the design parameters to be within $2\%$ of the beam target, $1\%$ of the depth target, and $\geq 99\%$ of the volume target, while also constraining the design with the forty-nine feasibility constraints to maintain design feasibility. The target draft is held constant, so $t*$ is scaled appropriately for each design at each generation during optimization.  This provides a buffer for the optimization algorithm to find low-drag designs around the test case targets. In this study, the two objective functions were the total resistance of a hull and the total resistance coefficient of a hull, which were evaluated using the total resistance coefficient regression model, $P_{C_T}$ at the test case's target speed. By leveraging the same regression model in both optimization and diffusion generation, we can directly compare the ability of each design method to produce low-drag designs. The optimization is expected to exploit the regression model, likely finding local minima and not a true minimum. This will be seen as a significant loss in accuracy when comparing predictions from the regression model to the original total resistance simulation. With the combined use of parallelized CPU computation (\textit{Intel Core i9-13900K}) and GPU (\textit{NVIDIA GeForce RTX 4090}), this optimization is performed in approximately 80 minutes per test case. 

\section{Results}
This section contains the results of the studies described in the Methods Section. The first subsection provides error measurements of diffusion-generated samples meeting the principal dimensions from the five test cases. The second subsection provides the results by generating low-resistance designs using the conditional diffusion model and the design optimization algorithm.

\subsection{Targeted Design Sampling with Conditional Diffusion Model}
For each design test case, 512 samples were generated with the full model, and 512 samples were generated with only feasibility guidance, $\nabla_{X_t}f_{\phi}(y|X_t)$. The feasibility rate and adherence to principal dimensions in each test case are provided in Table~\ref{table:tabGeneratedSamples}. 

\begin{table}[ht]
\begin{center}

\begin{tabular}{lcccccccc}
\hline
\multicolumn{1}{c}{\multirow{2}{*}{\textbf{Test Case}}} & \multicolumn{1}{c}{\multirow{2}{*}{\textbf{Model}}} & \multicolumn{1}{c}{\multirow{2}{*}{\textbf{Feasibility Rate}}} & \multicolumn{2}{c}{\textbf{Volume Error}} & \multicolumn{2}{c}{\textbf{Beam Error}} & \multicolumn{2}{c}{\textbf{Depth Error}} \\
\multicolumn{1}{c}{} & \multicolumn{1}{c}{} & \multicolumn{1}{c}{} & \multicolumn{1}{c}{\textit{Mean}} & \multicolumn{1}{c}{\textit{Std.}} & \multicolumn{1}{c}{\textit{Mean}} & \multicolumn{1}{c}{\textit{Std.}} & \multicolumn{1}{c}{\textit{Mean}} & \multicolumn{1}{c}{\textit{Std.}} \\
\hline
\multirow{2}{*}{Supercarrier} & Full Model & 67.77\% & -3.17\% & 8.83\% & 2.03\% & 2.98\% & -2.14\% & 1.82\% \\
                             & $\nabla_{X_t}f_{\phi}$ Only & 88.28\% & 2.53\% & 5.97\% & 1.46\% & 4.94\% & -1.65\% & 1.82\% \\
\hline
\multirow{2}{*}{Kayak} & Full Model & 86.91\% & 0.05\% & 4.52\% & -0.50\% & 2.75\% & -0.45\% & 1.77\% \\
 & $\nabla_{X_t}f_{\phi}$ Only & 95.70\% & 0.62\% & 3.50\% & 0.40\% & 3.55\% & 0.06\% & 1.63\% \\
 \hline
\multirow{2}{*}{NeoPanamax} & Full Model & 71.09\% & -2.89\% & 5.63\% & 1.41\% & 3.60\% & -0.44\% & 1.69\% \\
 & $\nabla_{X_t}f_{\phi}$ Only& 92.19\% & 2.26\% & 4.66\% & 0.18\% & 3.74\% & -0.01\% & 1.70\% \\
 \hline
\multirow{2}{*}{Frigate} & Full Model & 91.99\% & -0.87\% & 6.60\% & 0.22\% & 8.06\% & -0.31\% & 2.44\% \\
 & $\nabla_{X_t}f_{\phi}$ Only & 94.53\% & 0.83\% & 6.93\% & 1.65\% & 10.27\% & 0.19\% & 1.93\% \\
 \hline
\multirow{2}{*}{ROPAX ferry} & Full Model & 58.98\% & -16.73\% & 7.57\% & -4.74\% & 6.78\% & -1.02\% & 1.86\% \\
 & $\nabla_{X_t}f_{\phi}$ Only & 88.48\% & -2.80\% & 5.52\% & 0.38\% & 7.37\% & -0.34\% & 2.15\% \\
 \hline
\end{tabular}
\end{center}
\caption{This table provides the design feasibility rate and principal dimension errors relative to each test case for diffusion-generated samples using the full model and with feasibility guidance only.}
\label{table:tabGeneratedSamples}
\end{table}

The general trend among these test cases is that designs sampled with feasibility guidance only have higher feasibility rates and have tighter adherence to the test case's principal dimensions compared to hulls generated with the full model. This trend is seen in all five test cases. Generated hull designs within a 5\% error tolerance will be selected for further design analysis. This tolerance is a reasonable margin for large ship hull designs. A ship's displacement can easily vary by this much through changes in cargo, fuel, water, etc. Sampling with only feasibility guidance yields, on average, 58.3\% of hulls generated within a 5\% volume error across the five test cases. Sampling with the full model gives 37.0\% of total samples within a 5\% volume error tolerance. This measure was calculated with Equation~\ref{eq_Stats} for each design test case.

 \begin{equation}
  \eta_{E_{V_{t*}}} =\eta_{fease}\Phi\left(\frac{+5\% - \mu_{E_{V_{t*}}}}{\sigma_{E_{V_{t*}}}}\right) - \Phi\left(\frac{-5\% - \mu_{E_{V_{t*}}}} {\sigma_{E_{V_{t*}}}}\right)
\label{eq_Stats}
\end{equation}  
In the equation, $ \eta_{E_{V_{t*}}}$ is the percentage of samples within a 5\% volume error tolerance. This metric relies on the feasibility rate, $\eta_{fease}$, and the Gaussian cumulative distribution between +5\% and -5\% error given the mean ($\mu_{E_{V_{t*}}}$) and standard deviation($\sigma_{E_{V_{t*}}}$) of volume error.

\subsection{Drag Reduction with Guidance During Sampling}
For each design test case, hull designs were optimized using \textit{NSGA-II} to minimize the total resistance while satisfying the principal dimensions of the test case. After optimization, the total resistance of the 100 optimized hulls was calculated with the Michell Integral. The minimum total resistance calculated with the Michell Integral is listed in Table~\ref{table:tabRtSamples} for each test case. Similarly, the total resistance was calculated for the feasible hull designs among the 512 designs generated for each test case. The feasible hull designs were sorted into groups with volume errors less than 1\%,  5\%, and 10\% relative to the target volume for each design test case. Then, the number of hulls with a total resistance less than the optimized minimum total resistance was collected. The number of these low resistance samples within each volume error tolerance is listed in Table~\ref{table:tabRtSamples}. The final column in Table~\ref{table:tabRtSamples} lists the minimum total resistance among the diffusion-generated designs within a 5\% volume error tolerance for each design test case. These results will be further analyzed in the Discussion Section. 

\begin{table}[ht]
\begin{center}

\begin{tabular}{lccccccc}
\hline
\textbf{Test Case} & \textbf{\textit{NSGA-II} Min. $\mathbf{R_T}$} & \textbf{Model} & \multicolumn{3}{c}{\textbf{Number of Low $\mathbf{R_T}$ Hulls with $\mathbf{E_{V_{t*}}}$}} & \textbf{Sample Min. $\mathbf{R_T}$} &  $\Delta\mathbf{ R_T}$ \\

 &  \textit{[N]} &  & $\leq 1\%$ & $\leq 5\%$ & $\leq 10\%$ & \textit{[N]} & \\
 \hline
\multirow{2}{*}{Supercarrier} & \multirow{2}{*}{7,332,137.7} & Full Model & 0 & 5 & 11 & \multirow{2}{*}{4,883,089.3} & \multirow{2}{*}{-33.4\%}\\
 &  & $\nabla_{X_t}f_{\phi}$ Only& 0 & 1 & 1& \\
 \hline
\multirow{2}{*}{Kayak} & \multirow{2}{*}{11.18} & Full Model & 8 & 50 & 56 & \multirow{2}{*}{6.98} & \multirow{2}{*}{-37.6\%} \\
 &  & $\nabla_{X_t}f_{\phi}$ Only& 1 & 2 & 2& \\
 \hline
\multirow{2}{*}{NeoPanamax} & \multirow{2}{*}{3,931,834.1} & Full Model & 37 & 157 & 239 & \multirow{2}{*}{1,220,057.3} & \multirow{2}{*}{-69.0\%} \\
 &  & $\nabla_{X_t}f_{\phi}$ Only& 5 & 23 & 28& \\
 \hline
\multirow{2}{*}{Frigate} & \multirow{2}{*}{1,177,601.3} & Full Model & 30 & 130 & 178 & \multirow{2}{*}{874,617.0} & \multirow{2}{*}{-25.7\%} \\
 &  & $\nabla_{X_t}f_{\phi}$ Only & 26 & 87 & 114 &\\
 \hline
\multirow{2}{*}{ROPAX ferry} & \multirow{2}{*}{2,512,677.3} & Full Model & 1 & 9 & 41 & \multirow{2}{*}{206,537.0} & \multirow{2}{*}{-91.8\%} \\
 &  & $\nabla_{X_t}f_{\phi}$ Only & 42 & 208 & 307&\\
 \hline
\end{tabular}
\end{center}
\caption{This table lists the number of feasible, diffusion generated hull designs having a total resistance less than the minimum total resistance found through optimization. The number of hulls with low resistance increases as the volume error tolerance is increased. The final column lists the reduction in total resistance seen by a hull within a 5\% volume error generated by C-ShipGen.}
\label{table:tabRtSamples}
\end{table}

In general, as the volume error tolerance is loosened, more diffusion-sampled hull designs will have lower total resistance than the optimized hull design. This trend is seen in samples generated using the full model and among samples generated using only classifier guidance. Additionally, for four test cases, the full diffusion model produces low resistance designs within a 5\% volume error tolerance at a rate of 1.5x to 25x more frequently than without using performance guidance. In addition, the full diffusion model generated hulls with at least 25\% less total resistance than the \textit{NSGA-II} generated hulls while still aligning to the principal dimensions of each design test case. For the \textit{ROPAX} test case, the diffusion model with only classifier guidance created low-resistance designs with much higher success than the full diffusion model. Further analysis of the test cases can be found in the Discussion Section.

\begin{figure}[ht]%
\centering
\begin{minipage}[t]{.48\linewidth}
    \centering
    \includegraphics[width=3.25in]{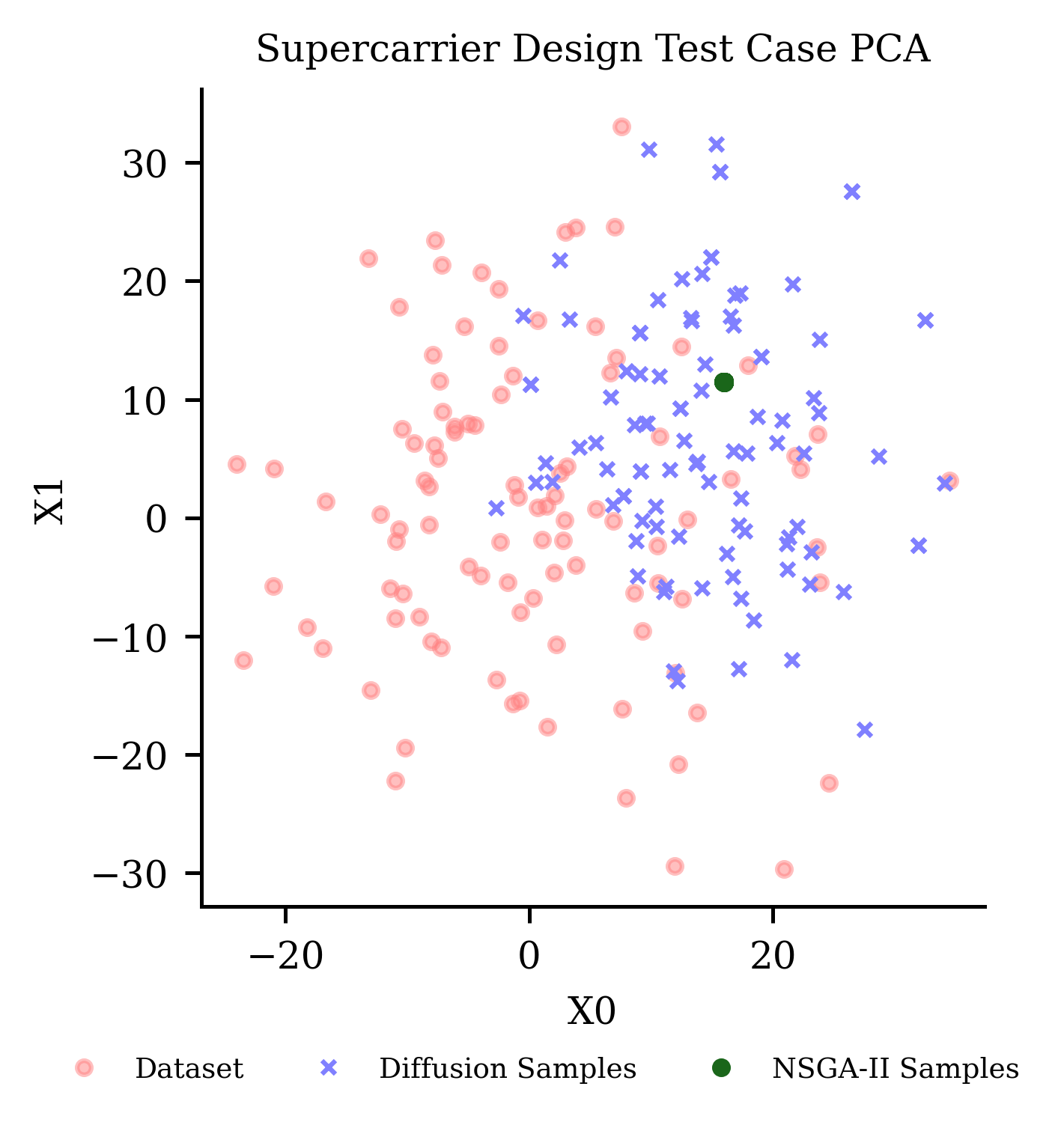}
    \captionof{figure}{Two-dimensional principal component analysis of the hull parameterization shows the relative distribution between dataset hull designs, diffusion-generated hull designs, and optimized hull designs for the supercarrier test case. The optimized hulls have much less design diversity than the diffusion-generated designs.}
    \label{fig:figure_Study0PCA} 
\end{minipage}\hfill%
\begin{minipage}[t]{.48\linewidth}
    \centering
    \includegraphics[width = 3.25in]{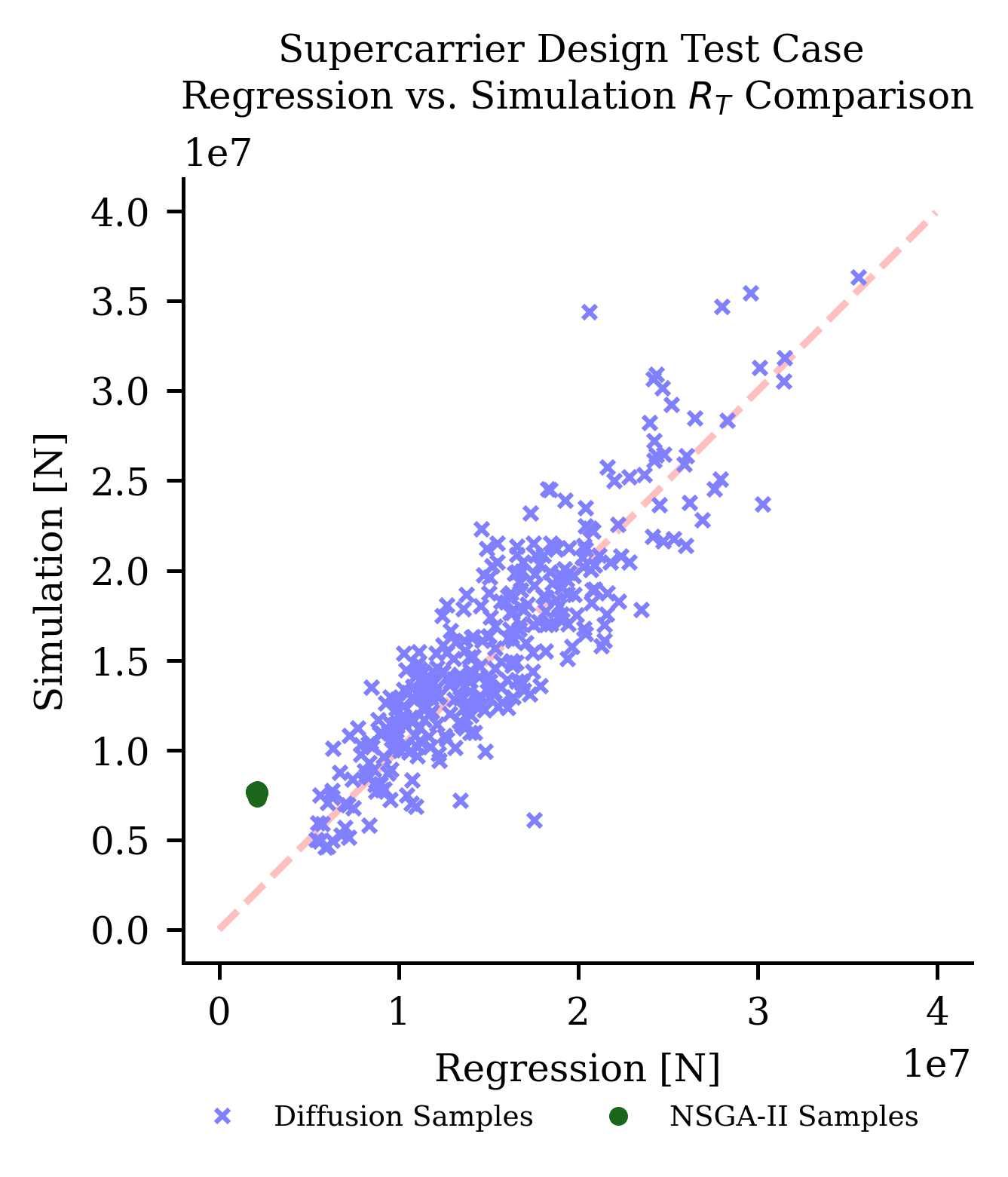}
    \captionof{figure}{Comparison of total resistance between simulation and regression for hull designs produced for the supercarrier design test case. The optimized designs have lower regression accuracy than the diffusion-generated designs.}
    \label{fig:figure_Study0_RT} 
\end{minipage}
\end{figure}

To illustrate the diversity of designs generated by the diffusion model, a two-dimensional principal component analysis (PCA) was performed with the design parameters from the training dataset, the diffusion-generated designs, and the \textit{NSGA-II} sampled designs. The PCA was fitted with the training dataset. Figure~\ref{fig:figure_Study0PCA} shows this PCA. The diffusion-generated samples maintain a high degree of diversity. This is expected as the diffusion model is trained to randomly generate designs that match the statistics of the training data. The optimized designs, however, are clustered around a single location on the PCA plot, suggesting these samples have low diversity. 

\begin{figure}[h]
\begin{center}
\setlength{\unitlength}{0.012500in}%
\includegraphics[width = 6.5in]{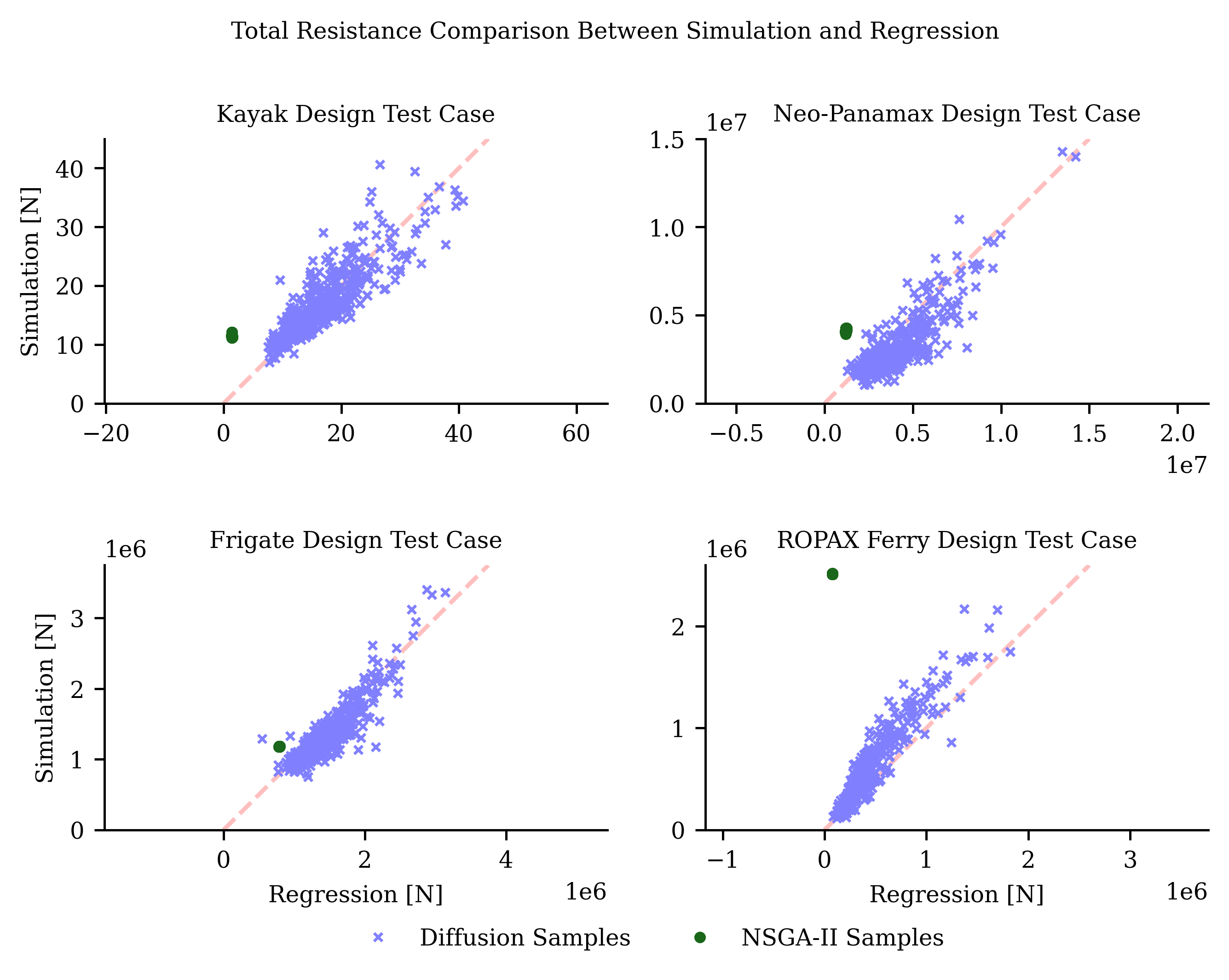}
\caption{Comparison of total resistance between simulation and regression for hull designs produced for the other design test cases. The total resistance of optimized designs is less accurately predicted by the regression model than the diffusion-generated designs.}
\label{fig:figure_testcases_RT} 
\end{center}
\end{figure}

After sampling, all feasible diffusion-generated designs and \textit{NSGA-II} generated designs were simulated with the total resistance simulation. This was done to compare the regression model accuracy to the simulation data.  Figure~\ref{fig:figure_Study0_RT} shows the total resistance of each hull using both regression and simulation plotted against each other for the supercarrier design test case. The red dashed line shows perfect regression, meaning that the regression predicts the exact simulation. The generated samples show high accuracy with the regression, except for a few outliers. This is expected as the diffusion-generated samples statistically resemble the training dataset. This dataset was used to train both the regression model and the diffusion model. Because diffusion-generated designs will statistically resemble the training data, they should have higher accuracy with the regression model. The hull designs created through optimization are significantly under-predicted compared to the total resistance calculated with the simulation. As mentioned in the methods section, this result was expected as the optimization algorithm exploited the regression model to find a minimum in the model that does exist in the simulation. This trend is seen across optimized hulls from the other four test cases, shown in Figure ~\ref{fig:figure_testcases_RT}.

\begin{figure}[ht]%
\centering

    \includegraphics[width=3.25in,height=3.25in]{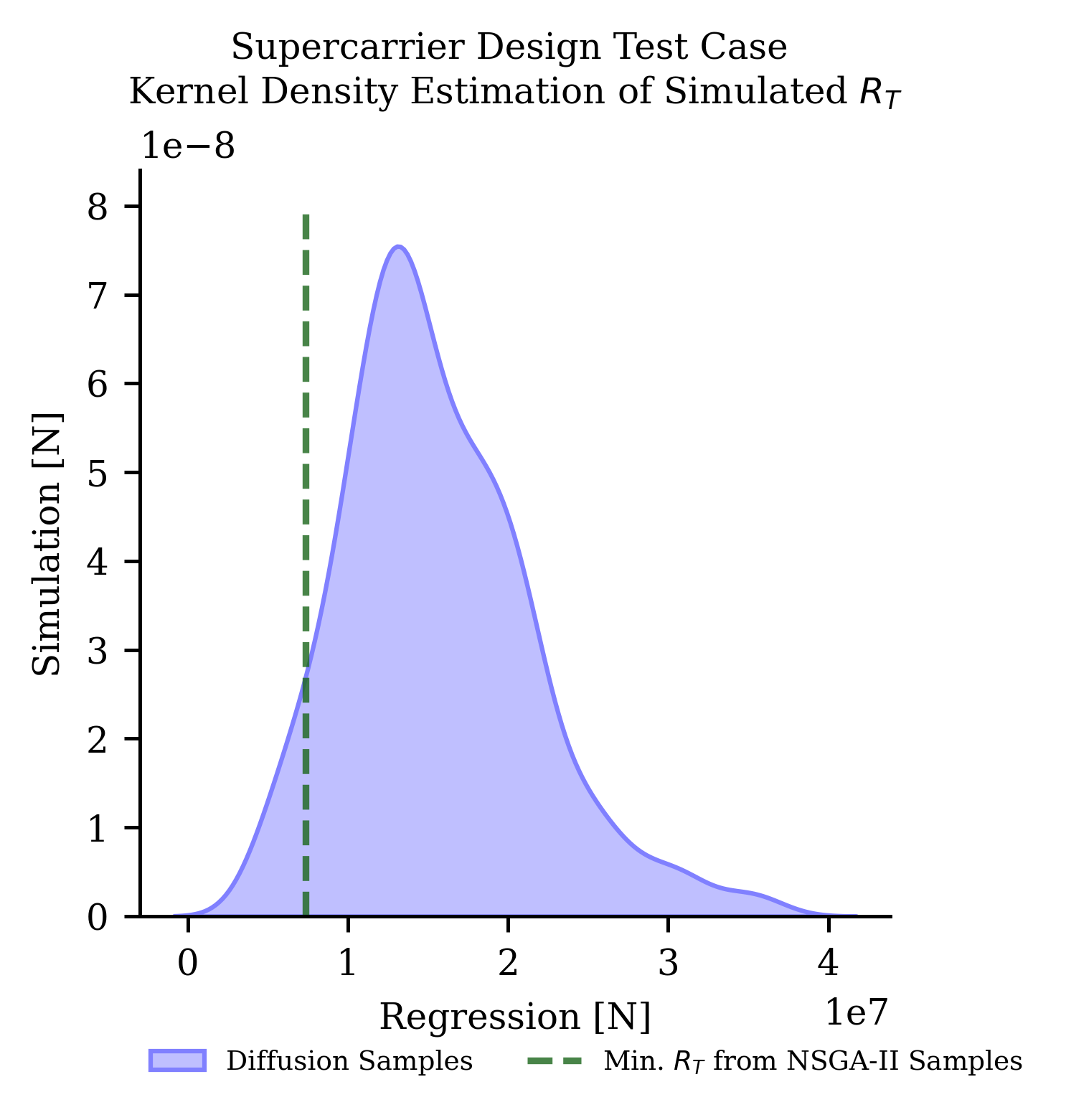}
    \captionof{figure}{Kernel Density Estimate (KDE) of the distribution of the simulated total resistance across the diffusion-generated samples for the supercarrier test case. The distribution shows that some of the generated samples have a total resistance less than the minimum total resistance found using the total resistance prediction as a surrogate model with \textit{NSGA-II}.}
    \label{fig:figure_Study0KDE} 
\end{figure}

\begin{figure}[h]
\begin{center}
\setlength{\unitlength}{0.012500in}%
\includegraphics[width = 6.5in]{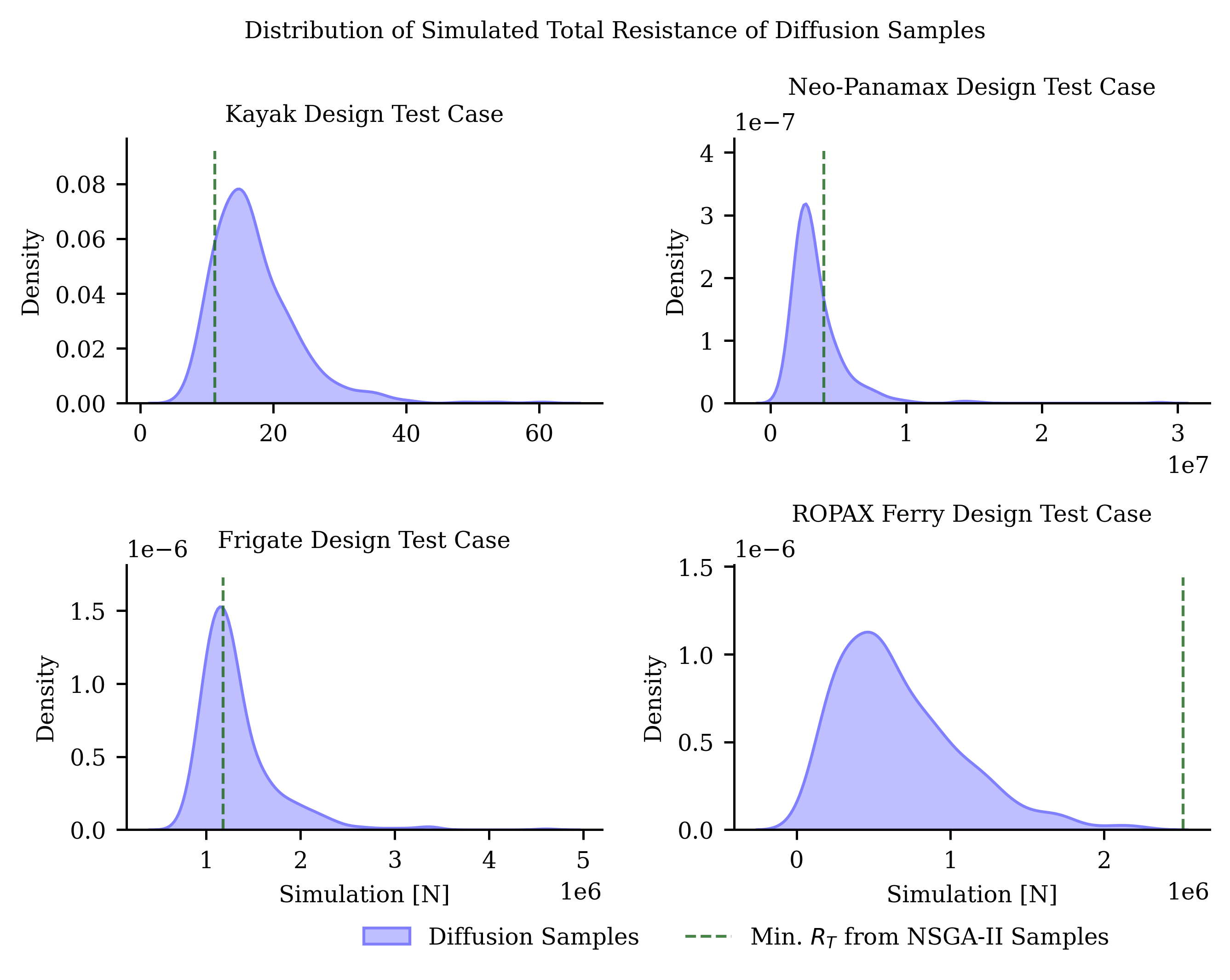}
\caption{KDE plots of the remaining four design test cases. The plots showcase the statistical distribution of total resistance among the diffusion-generated samples compared to the \textit{NSGA-II} generated samples. Depending on the test case, 15\% or more of the generated samples will have less total resistance than samples produced using \textit{NSGA-II} with the same surrogate model for total resistance prediction.}
\label{fig:figure_testcases_KDE} 
\end{center}
\end{figure}

To further visualize the resistance across the diffusion-generated samples, a kernel density estimate (KDE) of the distribution of the simulated total resistance is shown in Figure~\ref{fig:figure_Study0KDE}. Also included in the plot is the minimum simulated total resistance among hulls sampled with \textit{NSGA-II}. The KDE shows that approximately fifteen percent of diffusion-generated samples for this design test case will have lower total resistance than samples generated with \textit{NSGA-II} using the same surrogate model for drag prediction. This relative trend also appears with the other four test cases shown in Figure~\ref{fig:figure_testcases_KDE}. 

The optimized designs from the ROPAX ferry design test case exhibit the worst regression-simulation similarity among the test cases. The regression prediction for the optimized ROPAX ferry was off by nearly a factor of 10 compared to its simulation-calculated resistance. Further analysis of the error in simulation prediction is in the Results Section.

\begin{figure}[ht]
\begin{center}
\setlength{\unitlength}{0.012500in}%
\includegraphics[width = 6.5in]{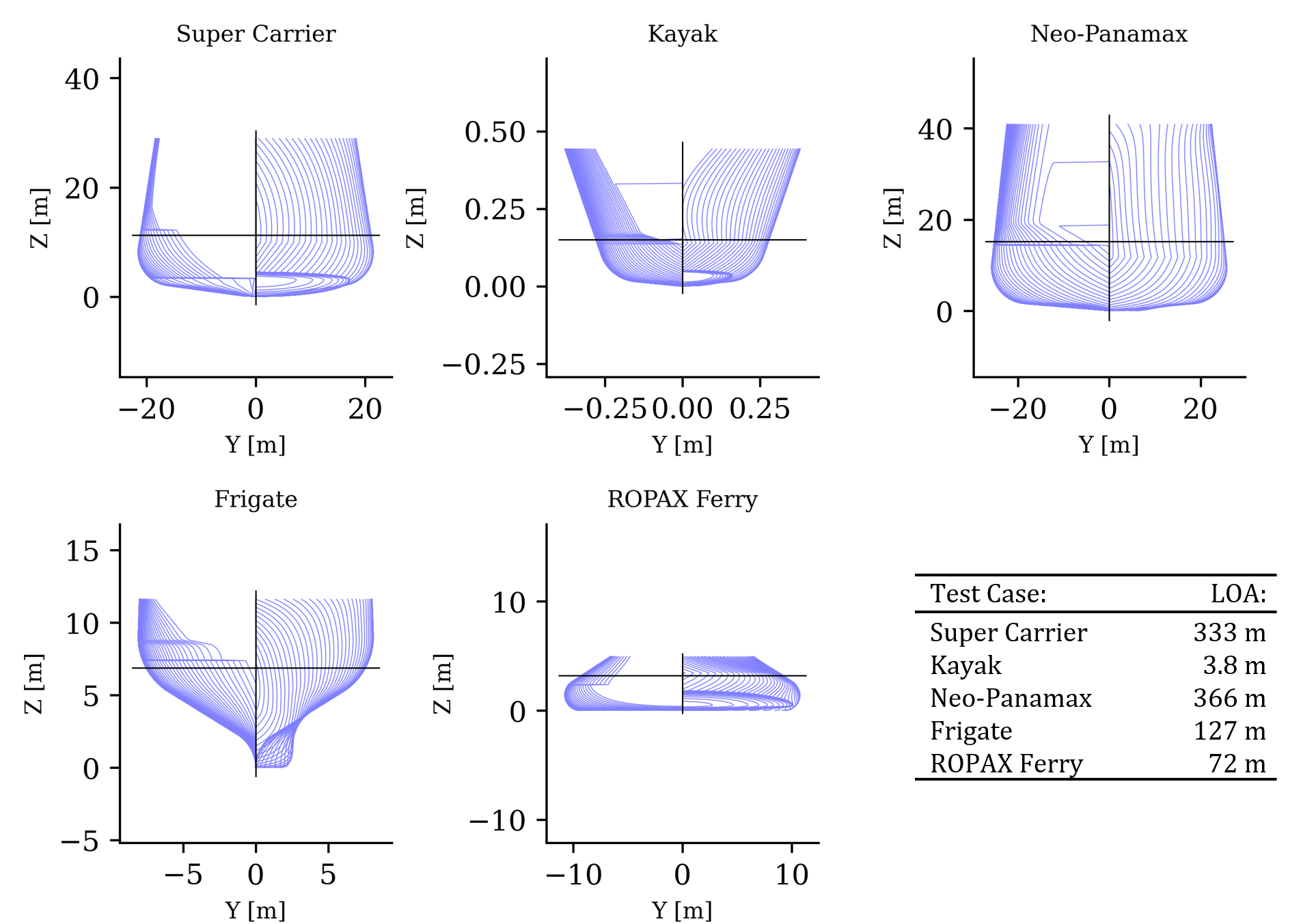}
\caption{The five hulls depicted are diffusion-generated hulls with the minimum total resistance within the 5\% volume error tolerance for each design test case. The cross-section drawings showcase station lines for the bow on the right side and the stern on the left side. Also included is the LOA for each design. }
\label{fig:figure_SampleHulls} 
\end{center}
\end{figure}

To exhibit some of the diffusion-generated hull designs, Figure~\ref{fig:figure_SampleHulls} showcases the station lines of hulls from each test case. The Figure showcases the hull design sampled among the 512 with the minimum resistance within the 5\% volume error tolerance for each design test case. The total resistance of these hulls is listed in Table~\ref{table:tabRtSamples}.

\section{Discussion}
This section provides a discussion of the results of the study. The first subsection analyzes the diffusion model's ability to satisfy the principal characteristics from the design test cases. The second subsection discusses the study on generating low-resistance designs. A third subsection discusses the ROPAX design case study compared to the other design test cases. The final subsection discusses the limitations of the C-ShipGen model in designing real-world hulls. 

\subsection{Targeted Sample Generation}
The study's findings underscore the effectiveness of the conditional diffusion model in generating feasible designs that closely adhere to design requirements. The tolerance to the user-defined principal characteristics decreases when performance guidance is implemented with the model during sampling. Performance guidance reduces the diffusion model's ability to generate hulls within a 5\% volume error tolerance by 36\%. The next subsection will discuss how performance guidance produces low-resistance hull designs more frequently while maintaining the 5\% volume error tolerance. 

While not every individual design meets the entirety of the specifications provided by a user during sampling, the diffusion model proves to be a computationally inexpensive tool for producing samples closely aligned with intended principal dimensions. This characteristic makes it particularly advantageous for early-stage design processes, where loosely following requirements allows for design exploration. The diffusion model produces many diverse designs for further in-depth design analysis. With an efficient down-selection process, many useful hull designs are quickly identified within a user-desired tolerance with C-ShipGen. 

\subsection{Optimization versus Diffusion Models for Design Generation}
Optimization represents a powerful approach to design exploration but has strong advantages and disadvantages. One key strength is the optimization's ability to generate samples within tighter tolerances to user-defined targets. Additionally, the optimization process excels at producing feasible designs with low resistance. On the other hand, design optimization for each test case can be slow and computationally expensive. For \textit{NSGA-II}, increasing the population size increases the time complexity of the algorithm by $O(N^2)$. Increasing the number of objectives increases its time complexity by $O(N)$~\cite{Deb2002NSGA2}. Optimization is limited by the diversity of samples and the computational complexity arising from increasing the number of samples. This inhibits design space exploration in early-stage design. Finally, as seen in the results, optimization exploits regression models, which leads to a loss in accuracy of the optimized design's prediction versus ground truth performance calculated with simulation. In this instance, the loss in accuracy of the total resistance regression surrogate model compared to the simulation leads to sub-optimal hull designs compared to those generated with the C-ShipGen model. 

The diffusion model presents a contrasting set of advantages and challenges. The computation to generate designs is significantly faster, allowing more design space exploration and discovery over-optimization methods. Increasing the number of generated samples increases the time complexity of C-ShipGen by $O(N)$. Increasing the number of objectives increases the time complexity of C-ShipGen by $O(N)$. Secondly, this diffusion model produces new designs without model retraining, which allows significant flexibility in its use. This flexibility permits hull design across all scales of real-world displacement hulls within a large range of typical operating speeds. This flexibility is thanks to the diversity of the samples in the training data. Thirdly, leveraging the total resistance regression model during the diffusion sampling process is particularly advantageous. The regression model and the diffusion model were trained with the same dataset. Since the diffusion model is trained to generate designs with statistical similarity to the training dataset, the regression model has high prediction accuracy for diffusion-generated designs. This similarity between the training data and sampled designs is why C-ShipGen saw a significant improvement in total resistance among generated samples compared to \textit{NSGA-II} using the same total resistance regression model. This trend was seen across the five design test cases in Figure~\ref{fig:figure_Study0_RT} and Figure~\ref{fig:figure_testcases_RT}. This accuracy did not hold in optimization-generated designs. 
In addition, the diffusion model better leveraged the regression model in sampling. This is shown through the proportion of designs in each test case with lower total resistance than the samples generated with \textit{NSGA-II} using the same regression model. Figure~\ref{fig:figure_Study0KDE} and Figure~\ref{fig:figure_testcases_KDE} show that depending on the test case, roughly fifteen percent or more of the diffusion-generated samples will have a lower total resistance than samples generated using \textit{NSGA-II}. However, the diffusion model has its own set of limitations. One major flaw is that the generated design needs to be sorted and filtered to identify low-resistance hull forms that meet the input conditioning within a desired tolerance. Additionally, while the diffusion model facilitates rapid design exploration, it does not provide a guarantee of finding an optimal solution, introducing an element of uncertainty into the design process.

Additionally, the diffusion model with performance guidance produces designs within a specified tolerance less frequently than the diffusion model without performance guidance, and even less frequently than optimization. Despite this compromise in tolerable design generation, these designs demonstrate a notable reduction in drag compared to their counterparts without performance guidance. Implementing performance guidance produces low resistance designs at a rate of 1.5x to 25x than the model without the implementation. The simultaneous benefits of lower drag and decreased feasibility highlight the nuanced impact of performance guidance on design outcomes. 

\subsection{Analysis on ROPAX Ferry Design}
Despite the general success of the other design test cases, the ROPAX ferry design proved difficult for the diffusion model, the regression model, and the optimization algorithm. This design test case was inspired by a real-world ferry operating in the State of Massachusetts in the United States. This particular hull has a significantly higher length-to-draft ratio, beam-to-draft ratio, and block coefficient compared to other design test cases. These comparisons are also true compared to hulls in the training data. As this design is different than most of the training data, the outcomes of the test case were expected to be poorer. This is particularly true with the total resistance prediction model. The regression model had completely inaccurate predictions of total resistance for the \textit{NSGA-II} generated samples. The diffusion-generated designs also had poor regression accuracy as well. A third consideration with this particular design that is outside the project's scope is that the resistance simulation itself is also highly inaccurate for this design. Michell's integral relies on the assumption that the hull is a slender body~\cite{michell1898Wave}. The ROPAX design is not necessarily slender compared to the other hulls. A new simulation method is needed to reasonably calculate the drag on a hull design like this one. 

\subsection{Limitations of Hull Design with C-ShipGen}
When designing any product with generative artificial intelligence, understanding the model's limitations will avoid negative consequences on real products designed with the model. C-ShipGen has several limitations. The first limitation of this model is the training data. C-ShipGen and other diffusion models generate designs statistically similar to the training data. Since C-ShipGen was trained on hulls that are not necessarily representative of real-world hull designs, C-ShipGen generates hulls that are not necessarily representative of real-world designs. The second limitation of C-ShipGen is the simulation used to generate the total resistance training data. While the Michell integral is not the most accurate simulation for real-world hull design, it was chosen to balance accuracy and computational cost. Creating training data with higher fidelity and more accurate simulations will give these models data with a better representation of real-world hull designs in water. In addition, leveraging more accurate simulations for creating training data will enhance claims of increased performance with generative design. The third limitation of C-ShipGen in its current implementation is that it only considers total resistance and volume displacement for design generation. Real ship hulls are designed considering other performance metrics such as seakeeping, stability, general arrangements, draft limitations, and countless other considerations for hull design. Therefore, hulls designed with C-ShipGen are not necessarily capable of performing safely in the real world without further analysis. These are some of the limitations of the current implementation of C-ShipGen to create hull designs with low total resistance. 
\section{Conclusion}
This work generated ship hulls with low resistance using a conditional diffusion model that considers the desired principal dimensions of the hull during design generation. This diffusion model is trained on a large set of nearly 83,000 diverse hull designs that allow for a comprehensive design space exploration with the model. In addition, a regression model was trained to predict the total resistance of a hull with variable speed and draft. The gradients of this regression model allowed the diffusion model to generate designs with low resistance. This regression model was also used as a surrogate model to optimize hulls while constraining the designs to user-defined principal dimensions and design speeds. The optimization study was performed using \textit{NSGA-II}. Five design test cases demonstrated the ability of C-ShipGen to generate hull designs across all scales of displacement hulls and many different dimensional properties found in different ship classes. Additionally, C-ShipGen was able to generate designs with greater diversity than \textit{NSGA-II}, while creating designs with better predictive alignment between the regression model and the simulation used in the training data. A proportion of the diffusion-generated designs in each test also had a total resistance less than the samples generated with \textit{NSGA-II}. In all five test cases, C-ShipGen produced hull designs with at least 25\% less total resistance than \textit{NSGA-II} generated samples. An additional advantage of the diffusion model is that the diversity of these designs allows for efficient design space exploration in early-stage design. 

Creating hull designs with reduced resistance will reduce the need to fuel ships, reducing the cost to operate the ship and reducing its emissions. Future work with generative artificial intelligence for ship design will continue to explore the systems-level design of ships. Training models to explore the nuanced complexity of designing a ship system can yield better efficiencies and reduce costs for the marine industry. Through this work, the economic prospect of leveraging generative artificial intelligence to design ship hulls is demonstrated by C-ShipGen.

\section{Acknowledgements}
This research is funded by the United States Department of Defense, Office of Naval Research, via the National Defense Science and Engineering Graduate (NDSEG) Fellowship program. The authors would like to thank MIT Supercloud for providing some of the computational resources needed to perform this work~\cite{supercloud}. Data, Code, and Trained models for C-ShipGen will be available by the time the time of publication at \url{https://decode.mit.edu/projects/C_ShipGen/}.

\printbibliography

@article{yang2023product,
  title={A product form design method integrating Kansei engineering and diffusion model},
  author={Yang, Chaoxiang and Liu, Fei and Ye, Junnan},
  journal={Advanced Engineering Informatics},
  volume={57},
  pages={102058},
  year={2023},
  publisher={Elsevier}
}

@article{ploennigs2023diffusion,
  title={Diffusion models for computational design at the example of floor plans},
  author={Ploennigs, Joern and Berger, Markus},
  journal={arXiv preprint arXiv:2307.02511},
  year={2023}
}

@article{lew2023single,
  title={Single-shot forward and inverse hierarchical architected materials design for nonlinear mechanical properties using an attention-diffusion model},
  author={Lew, Andrew J and Buehler, Markus J},
  journal={Materials Today},
  volume={64},
  pages={10--20},
  year={2023},
  publisher={Elsevier}
}

@ARTICLE{Deb2002NSGA2,
  author={Deb, K. and Pratap, A. and Agarwal, S. and Meyarivan, T.},
  journal={IEEE Transactions on Evolutionary Computation}, 
  title={A fast and elitist multiobjective genetic algorithm: NSGA-II}, 
  year={2002},
  volume={6},
  number={2},
  pages={182-197},
  doi={10.1109/4235.996017}}

@article{bagazinski2023shipgen,
  title={ShipGen: A Diffusion Model for Parametric Ship Hull Generation with Multiple Objectives and Constraints},
  author={Bagazinski, Noah J and Ahmed, Faez},
  journal={Journal of Marine Science and Engineering},
  volume={11},
  number={12},
  pages={2215},
  year={2023},
  publisher={MDPI}
}

@article{yonekura2023designing,
  title={Designing ship hull forms using generative adversarial networks},
  author={Yonekura, Kazuo and Omori, Kotaro and Qi, Xinran and Suzuki, Katsuyuki},
  journal={arXiv preprint arXiv:2311.05470},
  year={2023}
}

@article{evans1959basic,
  title={Basic design concepts},
  author={Evans, J Harvey},
  journal={Journal of the American Society for Naval Engineers},
  volume={71},
  number={4},
  pages={671--678},
  year={1959},
  publisher={Wiley Online Library}
}

@article{wang2022shipEncoding,
  title={Three-dimensional ship hull encoding and optimization via deep neural networks},
  author={Wang, Yuyang and Joseph, Joe and Aniruddhan Unni, TP and Yamakawa, Soji and Barati Farimani, Amir and Shimada, Kenji},
  journal={Journal of Mechanical Design},
  volume={144},
  number={10},
  pages={101701},
  year={2022},
  publisher={American Society of Mechanical Engineers}
}

@article{maze2022topodiff,
title={Diffusion Models Beat GANs on Topology Optimization}, volume={37}, url={https://ojs.aaai.org/index.php/AAAI/article/view/26093}, DOI={10.1609/aaai.v37i8.26093},  number={8}, journal={Proceedings of the AAAI Conference on Artificial Intelligence}, author={Mazé, François and Ahmed, Faez}, year={2023}, month={06}, pages={9108-9116} }

@article{lin2017feature,
  title={Feature-based estimation of preliminary costs in shipbuilding},
  author={Lin, Cheng-Kuan and Shaw, Heiu-Jou},
  journal={Ocean Engineering},
  volume={144},
  pages={305--319},
  year={2017},
  publisher={Elsevier}
}

@article{ao2021artificial,
  title={An artificial intelligence-aided design (AIAD) of ship hull structures},
  author={Ao, Yu and Li, Yunbo and Gong, Jiaye and Li, Shaofan},
  journal={Journal of Ocean Engineering and Science},
  year={2021},
  publisher={Elsevier}
}

@article{ao2022artificial,
  title={Artificial Intelligence Design for Ship Structures: A Variant Multiple-Input Neural Network-Based Ship Resistance Prediction},
  author={Ao, Yu and Li, Yunbo and Gong, Jiaye and Li, Shaofan},
  journal={Journal of Mechanical Design},
  volume={144},
  number={9},
  pages={091707},
  year={2022},
  publisher={American Society of Mechanical Engineers}
}

@article{khan2022geometric,
  title={Geometric moment-dependent global sensitivity analysis without simulation data: application to ship hull form optimisation},
  author={Khan, Shahroz and Kaklis, Panagiotis and Serani, Andrea and Diez, Matteo},
  journal={Computer-Aided Design},
  volume={151},
  pages={103339},
  year={2022},
  publisher={Elsevier}
}

@article{khan2022shape,
  title={Shape-supervised dimension reduction: Extracting geometry and physics associated features with geometric moments},
  author={Khan, Shahroz and Kaklis, Panagiotis and Serani, Andrea and Diez, Matteo and Kostas, Konstantinos},
  journal={Computer-Aided Design},
  volume={150},
  pages={103327},
  year={2022},
  publisher={Elsevier}
}

@article{peri2001design,
  title={Design optimization of ship hulls via CFD techniques},
  author={Peri, Daniele and Rossetti, Michele and Campana, Emilio F},
  journal={Journal of ship research},
  volume={45},
  number={02},
  pages={140--149},
  year={2001},
  publisher={SNAME}
}

@article{brown2003multiple,
  title={Multiple-objective optimization in naval ship design},
  author={Brown, Alan and Salcedo, Juan},
  journal={Naval Engineers Journal},
  volume={115},
  number={4},
  pages={49--62},
  year={2003},
  publisher={Wiley Online Library}
}

@book{newman2018marine,
  title={Marine hydrodynamics},
  author={Newman, John Nicholas},
  year={2018},
  publisher={The MIT press}
}

@book{read2009drag,
  title={A drag estimate for concept-stage ship design optimization},
  author={Read, Douglas},
  year={2009},
  publisher={The University of Maine}
}

@inproceedings{zhang2018parametric,
  title={Parametric Method Using Grasshopper for Bulbous Bow Generation},
  author={Zhang, Yongxing and Kim, Dong-Joon and Bahatmaka, Aldias},
  booktitle={2018 International Conference on Computing, Electronics \& Communications Engineering (iCCECE)},
  pages={307--310},
  year={2018},
  organization={IEEE}
}

@article{chrismianto2014parametric,
  title={Parametric bulbous bow design using the cubic Bezier curve and curve-plane intersection method for the minimization of ship resistance in CFD},
  author={Chrismianto, Deddy and Kim, Dong-Joon},
  journal={Journal of Marine Science and Technology},
  volume={19},
  pages={479--492},
  year={2014},
  publisher={Springer}
}

@article{michell1898Wave,
  title={XI. The wave-resistance of a ship},
  author={Michell, John Henry},
  journal={The London, Edinburgh, and Dublin Philosophical Magazine and Journal of Science},
  volume={45},
  number={272},
  pages={106--123},
  year={1898},
  publisher={Taylor \& Francis}
}

@article{lu2016hydrodynamic,
  title={A hydrodynamic optimization design methodology for a ship bulbous bow under multiple operating conditions},
  author={Lu, Yu and Chang, Xin and Hu, An-kang},
  journal={Engineering Applications of Computational Fluid Mechanics},
  volume={10},
  number={1},
  pages={330--345},
  year={2016},
  publisher={Taylor \& Francis}
}

@article{demo2021hull,
  title={Hull shape design optimization with parameter space and model reductions, and self-learning mesh morphing},
  author={Demo, Nicola and Tezzele, Marco and Mola, Andrea and Rozza, Gianluigi},
  journal={Journal of Marine Science and Engineering},
  volume={9},
  number={2},
  pages={185},
  year={2021},
  publisher={Multidisciplinary Digital Publishing Institute}
}

@inproceedings{marlantes2021Modeling,
    title = {Modeling Vertical Planing Boat Motions using a Neural-Corrector Method},
    author ={Marlantes, Kyle and Maki, Kevin},
    volume = {Day 1 Tue, October 26, 2021},
    booktitle = {SNAME International Conference on Fast Sea Transportation},
    year = {2021},
    month = {10},
    doi = {10.5957/FAST-2021-014},
    url = {https://doi.org/10.5957/FAST-2021-014},
   
    eprint = {https://onepetro.org/snamefast/proceedings-pdf/FAST21/1-FAST21/D011S002R005/2520674/sname-fast-2021-014.pdf},
}

@article{silva2023implementation,
  title={Implementation of the Critical Wave Groups Method with Computational Fluid Dynamics and Neural Networks},
  author={Silva, Kevin M and Maki, Kevin J},
  journal={arXiv preprint arXiv:2301.09834},
  year={2023}
}

@article{abbas2023deepmorpher,
  title={DeepMorpher: deep learning-based design space dimensionality reduction for shape optimisation},
  author={Abbas, Asad and Rafiee, Ashkan and Haase, Max},
  journal={Journal of Engineering Design},
  volume={34},
  number={3},
  pages={254--270},
  year={2023},
  publisher={Taylor \& Francis}
}

@book{AppNavArch,
	author	  =	"R. Zubaly",
	title	  =	"Applied Naval Architecture",
	publisher
=	"Cornell Maritime Press",
	year	 =	1996,
ISBN = 9780870334757}

@article{HullOpt,
	author	=	"Y. Feng and O. el Moctar and T.E. Schellin",
	title	=	"Parametric Hull Form Optimization of Containerships for Minimum Resistance in Calm Water and in Waves",
	journal =   "Journal of Marine Science and Applications",
	month   =  {01},
	year	=	2022}

@article{PSOShip_Opt,
    author = {Knight, Joshua T.  and Zahradka, Frank T. and Singer, David J.  and Collette, Matthew D.},
    title = "{Multiobjective Particle Swarm Optimization of a Planing Craft with Uncertainty}",
    journal = {Journal of Ship Production and Design},
    volume = {30},
    number = {04},
    pages = {194-200},
    year = {2014},
    month = {11},
    issn = {2158-2866},
    doi = {10.5957/jspd.2014.30.4.194},
    url = {https://doi.org/10.5957/jspd.2014.30.4.194},
    eprint = {https://onepetro.org/JSPD/article-pdf/30/04/194/2204981/sname-jspd-2014-30-4-194.pdf},
}

@article{PSO_Multi_Opt,
    author = {Knight, Joshua T.  and Singer, David J.  and Collette, Matthew D.},
    title = "{Testing of a spreading mechanism to promote diversity in multi-objective particle swarm optimization}",
    journal = {Optimization and Engineering},
    volume = {16},
    pages = {279-302},
    year = {2015},
    month = {06}
}

@INPROCEEDINGS{supercloud,
  author={Reuther, Albert and Kepner, Jeremy and Byun, Chansup and Samsi, Siddharth and Arcand, William and Bestor, David and Bergeron, Bill and Gadepally, Vijay and Houle, Michael and Hubbell, Matthew and Jones, Michael and Klein, Anna and Milechin, Lauren and Mullen, Julia and Prout, Andrew and Rosa, Antonio and Yee, Charles and Michaleas, Peter},
  booktitle={2018 IEEE High Performance extreme Computing Conference (HPEC)}, 
  title={Interactive Supercomputing on 40,000 Cores for Machine Learning and Data Analysis}, 
  year={2018},
  volume={},
  number={},
  pages={1-6},
  doi={10.1109/HPEC.2018.8547629}}

@inproceedings{bagazinski2023shipD,
  title={Ship-D: Ship Hull Dataset for Design Optimization using Machine Learning},
  author={Bagazinski, Noah J and Ahmed, Faez},
  booktitle={International Design Engineering Technical Conferences and Computers and Information in Engineering Conference},
  year={2023},
  organization={American Society of Mechanical Engineers}
}

@article{ho2020denoising,
  title={Denoising diffusion probabilistic models},
  author={Ho, Jonathan and Jain, Ajay and Abbeel, Pieter},
  journal={Advances in neural information processing systems},
  volume={33},
  pages={6840--6851},
  year={2020}
}

@article{dhariwal2021diffusion,
  title={Diffusion models beat gans on image synthesis},
  author={Dhariwal, Prafulla and Nichol, Alexander},
  journal={Advances in neural information processing systems},
  volume={34},
  pages={8780--8794},
  year={2021}
}

@article{ramesh2022hierarchical,
  title={Hierarchical text-conditional image generation with clip latents},
  author={Ramesh, Aditya and Dhariwal, Prafulla and Nichol, Alex and Chu, Casey and Chen, Mark},
  journal={arXiv preprint arXiv:2204.06125},
  year={2022}
}

@inproceedings{rombach2021highresolution,
  title={High-Resolution Image Synthesis with Latent Diffusion Models},
  author={Rombach, Robin and Blattmann, Andreas and Lorenz, Dominik and Esser, Patrick and Ommer, Bj{\"o}rn},
  booktitle={2022 IEEE/CVF Conference on Computer Vision and Pattern Recognition (CVPR)},
  pages={10674--10685},
  year={2022},
  organization={IEEE}
}

@article{khan2023shiphullgan,
  title={ShipHullGAN: A generic parametric modeller for ship hull design using deep convolutional generative model},
  author={Khan, Shahroz and Goucher-Lambert, Kosa and Kostas, Konstantinos and Kaklis, Panagiotis},
  journal={Computer Methods in Applied Mechanics and Engineering},
  volume={411},
  pages={116051},
  year={2023},
  publisher={Elsevier}
}

@phdthesis{shaeffer2023application,
  title={Application of Artificial Neural Networks to Early-Stage Hull Form Design},
  author={Shaeffer, Austin},
  year={2023},
  school={George Mason University}
}

@article{hodgesai,
    author = {Hodges, Justin and Wheeler, M and Belhocine, M and Henry, J},
    year = {2022},
    month = {09},
    pages = {},
    title = {AI/ML APPLICATIONS FOR SHIP DESIGN},
    journal = {ICCAS 2022},
    doi = {10.3940/rina.iccas.2022.46}
}

@article{arechiga2023drag,
  title={Drag-guided diffusion models for vehicle image generation},
  author={Arechiga, Nikos and Permenter, Frank and Song, Binyang and Yuan, Chenyang},
  journal={arXiv preprint arXiv:2306.09935},
  year={2023}
}

@article{giannone2023learning,
  title={Learning from Invalid Data: On Constraint Satisfaction in Generative Models},
  author={Giannone, Giorgio and Regenwetter, Lyle and Srivastava, Akash and Gutfreund, Dan and Ahmed, Faez},
  journal={arXiv preprint arXiv:2306.15166},
  year={2023}
}

@inproceedings{shaeffer2020application,
  title={Application of Machine Learning to Early-Stage Hull Form Design},
  author={Shaeffer, Austin Kyle and Wilson, Wesley and Yang, Chi},
  booktitle={SNAME Maritime Convention},
  pages={D043S019R002},
  year={2020},
  organization={SNAME}
}

@article{giannone2023aligning,
  title={Aligning Optimization Trajectories with Diffusion Models for Constrained Design Generation},
  author={Giannone, Giorgio and Srivastava, Akash and Winther, Ole and Ahmed, Faez},
  journal={arXiv preprint arXiv:2305.18470},
  year={2023}
}

@misc{liu2023zero1to3,
  title={Zero-1-to-3: Zero-shot one image to 3d object},
  author={Liu, Ruoshi and Wu, Rundi and Van Hoorick, Basile and Tokmakov, Pavel and Zakharov, Sergey and Vondrick, Carl},
  booktitle={Proceedings of the IEEE/CVF International Conference on Computer Vision},
  pages={9298--9309},
  year={2023}
}

\end{document}